# Speed of sound and phase equilibria for (CO$_2$ + C$_3$H$_8$) mixtures

Daniel Lozano-Martín[a]; Rodrigo Susial[a], Pedro Hernández[a,b] Teresa E. Fernández-Vicente[c], M. Carmen Martín[a], José J. Segovia[a,*]

[a] TERMOCAL Research Group, Research Institute on Bioeconomy, Escuela de Ingenierías Industriales, Universidad de Valladolid, Paseo del Cauce, 59, 47011 Valladolid, Spain.

[b] Instituto Nacional de Técnica Aeroespacial (INTA), 28850 Torrejón de Ardoz, Madrid, Spain

[c] Centro Español de Metrología (CEM), Alfar 2, 2876 Tres Cantos, Madrid, Spain.

* Corresponding author e-mail: jose.segovia@eii.uva.es

**Abstract**

This work presents phase envelope and speed of sound data for the (0.60 CO$_2$ + 0.40 C$_3$H$_8$) and (0.80 CO$_2$ + 0.20 C$_3$H$_8$) binary mixtures. Phase equilibria was measured using a cylindrical resonator working in the microwave band whereas an acoustic resonator was used for speed of sound measurements. The experimental results were compared GERG-2008 equation of state, obtaining average absolute deviations by 0.24% in pressure for phase equilibria data and 0.025% for speed of sound data. Speed of sound values were used to derive perfect-gas heat capacities, acoustic virial coefficients and second density virial coefficients. In addition, AGA8-DC92 EoS performance was checked for the results derived from speeds of sound.

# 1. Introduction.

Several uses have been analyzed for (carbon dioxide + propane) mixtures. Some authors have proposed and studied these mixtures as natural mixtures with low global warming and zero ozone depletion potentials for substitution of classical refrigerants [1,2]. Other researches have explored the application of these mixtures as non-toxic and non-flammable fluid for power generation by organic Rankine cycles of low-grade heat [3,4]. This binary system is also used as solvent in supercritical phase for the extraction of oils from seeds and for the hydrogenation of vegetable oils and fats, due to the improvement in the mass transfer rates and reduction in the amount of reactant required and undesirable by-products [5–8]. Furthermore, it has attracted interest as working fluid for enhanced oil recovery methods by its injection into specific heavy oil reservoirs [9]. All studies agree that accurate knowledge of the thermophysical properties for these mixtures is required.

The reference thermodynamic model for these mixtures is the GERG-2008 equation of state (EoS) [10,11], originally developed for natural gas-like mixtures. The application range is limited to mole fractions $x_{CO_2} \leq 0.30$ and $x_{C_3H_8} \leq 0.14$ in the intermediate quality range, covering the pressure range $p = (0 - 35)$ MPa and temperature range $T = (90 - 450)$ K in the normal range. The binary interaction between carbon dioxide and propane is fitted to selected $(p, \rho, T)$, saturated liquid density and vapor-liquid equilibrium data, but the quality and extension of the experimental data sets is not sufficient for the development of a departure function, thus the interactions are only adjusted through the reducing functions. Apart from the recommendation of the development of a new specific departure function for mixtures of $(CO_2 + C_3H_8)$ when comprehensive and accurate data are available, GERG-2008 report [10,11] indicates also the necessity of further measurements for improving the data situation for caloric properties for this system. This is because the only speed of sound measurements founded in the literatures are those of Lin et al. [12], restricted to mole fractions $x_{C_3H_8} \leq 0.938$. The covered range in composition is constrained to a very low range in the liquid and supercritical phase because the goal at this research is the use of propane as a doping impurity for the catalytic inhibition of the vibrational relaxation phenomena that causes a large sound absorption and dispersion in pure carbon dioxide, which prevents the accurate study of the speed of sound with typical acoustic cavities at high frequency.

Thus, the present work focuses on the study of the speed of sound and phase equilibria of two mixtures of nominal molar compositions (0.60 $CO_2$ + 0.40 $C_3H_8$) and (0.80 $CO_2$ + 0.20 $C_3H_8$). These are mixtures with a composition around a mole fraction $x_{C3H8} = 0.3$ since, at this composition, the mixture reaches the limit of flammability and is usually chosen as optimum for the afore-mentioned applications.

Speed of sound measurements are performed with one of the state-of-art techniques, the spherical acoustic resonator, and the results are compared to the GERG-2008 EoS [10,11] and to AGA8-DC92 EoS [13,14] which is already widely used in the industry in order to check the uncertainty of these models. In addition, the perfect-gas isochoric and isobaric heat capacities, along with the acoustic virial coefficients are derived from the speed of sound data to assist in the development of future reliable correlations for improving the thermodynamic models. The density second virial coefficients have been determined from the acoustic ones to provide a robust interpretation of the speed of sound measurements and further comparison with the literature. Finally, phase equilibria were also determined using a microwave cylindrical resonator, recently developed in our laboratory [15].

## 2. Materials and Methods.

### 2.1 Mixtures.

Two binary ($CO_2$ + $C_3H_8$) mixtures were prepared at the Spanish Center of Metrology (CEM) in Madrid, Spain, according to the standard gravimetric procedure described in the ISO 6142-1 [16] and ISO 6143 [17]. Supplier, purity, molar mass and critical parameters of the pure gases used in the preparation are reported in Table 1, which also contains composition and corresponding expanded ($k$ = 2) uncertainty for the two mixtures. Carbon dioxide and propane were used without further purification, but their impurities were considered in the mixture preparation and taken into account in the reported uncertainty of the composition. The mixtures were homogenized by heating and rolling and checked by gas chromatography (GC), with differences between gravimetric composition and GC analysis within uncertainties.

**Table 1.** Purity, supplier, critical parameters of the pure components used for preparation of the binary ($CO_2$ + $C_3H_8$) mixtures at CEM, molar composition $x_i$ and expanded ($k = 2$) uncertainty $U(x_i)$ of the binary ($CO_2$ + $C_3H_8$) mixtures studied in this work.

| | Supplier | Purity / mol fraction[b] | CAS number | $M$ / g·mol$^{-1}$ | Critical parameters[a] | |
|---|---|---|---|---|---|---|
| | | | | | $T_c$ / K | $p_c$ / MPa |
| Carbon Dioxide | Air Liquide | ≥ 0.99998 | 124-38-9 | 44.010 | 304.13 | 7.3773 |
| Propane | Air Liquide | ≥ 0.99950 | 74-98-6 | 44.096 | 369.89 | 4.2512 |

| | (0.60 $CO_2$ + 0.40 $C_3H_8$) | | | (0.80 $CO_2$ + 0.20 $C_3H_8$) | | |
|---|---|---|---|---|---|---|
| Components | $10^2 \cdot x_i$ / mol/mol | $10^2 \cdot U(x_i)$ / mol/mol | Impurities | $10^2 \cdot x_i$ / mol/mol | $10^2 \cdot U(x_i)$ / mol/mol | Impurities |
| Carbon Dioxide | 60.059 | 0.029 | ≤ 80 parts in $10^6$ of other hydrocarbons | 80.110 | 0.015 | ≤ 40 parts in $10^6$ of other hydrocarbons |
| Propane | 39.941 | 0.021 | | 19.890 | 0.011 | |

[a] Critical parameters were obtained by using the default equation for each substance in REFPROP software [18], the reference equation of state for carbon dioxide [19] and the reference equation of state for propane [20].

[b] The stated purities were provided by the suppliers.

## 2.2 Experimental techniques.

### 2.2.1 Acoustic spherical resonator.

A schematic diagram of the experimental apparatus, for measuring speeds of sound, is depicted in Figure 1.

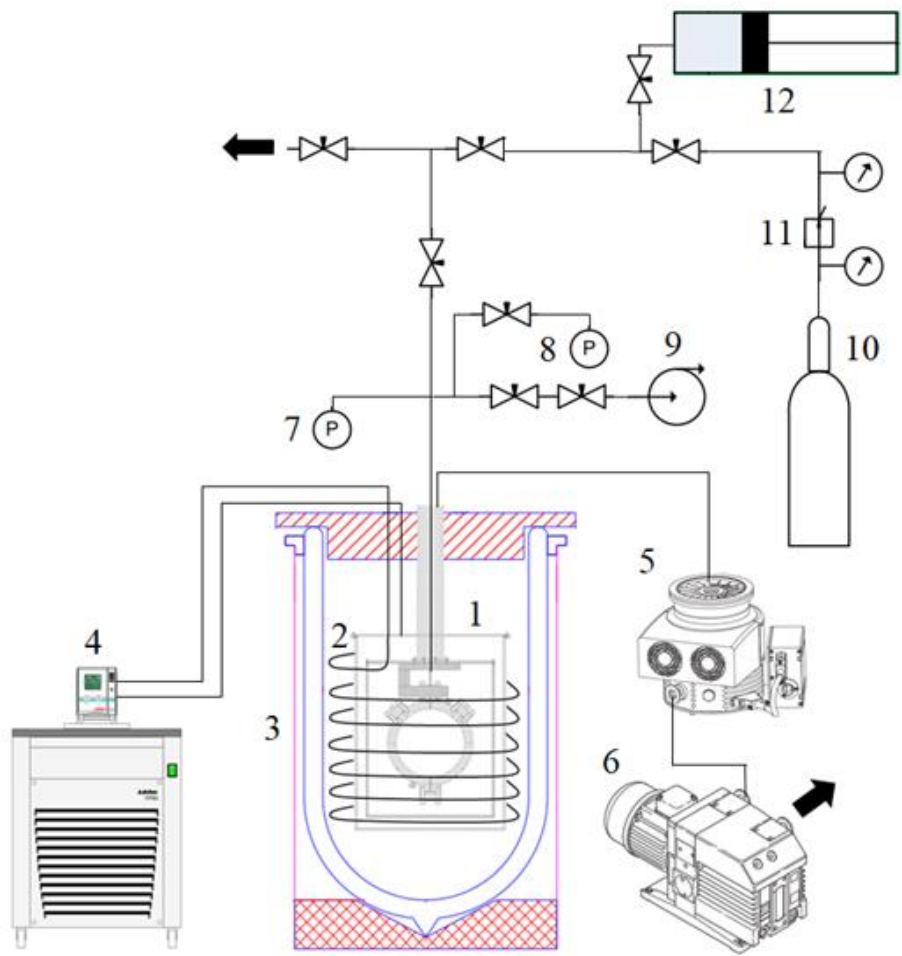

**Figure 1.** Schematic diagram of the technique for measuring speeds of sound: 1. Acoustic cavity; 2. Cooling coil; 3. Ethanol Dewar; 4. Cooling bath; 5. Turbomolecular pump; 6. Centrifugal pump; 7. pressure transducer (up to 20 MPa); 8. pressure transducer (up to 2 MPa); 9. Vacuum pump; 10. Gas sample bottle; 11. Pressure regulator; 12. Hand operated piston pump.

The setup comprises the spherical acoustic resonator of 40 mm internal radius in A321-type stainless-steel, designed to operate at pressures and temperatures up to 20 MPa and 475 K, which serves as acoustic cavity and pressure-tight vessel. An internal and an external stainless-steel shell, provided with a shield of aluminum over fibber-glass and under vacuum inside, prevent heat losses by radiation and convection. A cooled ethanol Dewar fed by a stirred thermal bath (FP89, Julabo) is used to maintain the system around ten degrees below the desire temperature and the cavity is heated by conduction using several electrical heaters around the internal shell. A gas manifold was designed to fill and evacuate the resonator with the gas sample from the bottle using a hand-operated gas piston and a vacuum pump (Trivac D8B, Leybold). More details about the construction and calibration of the internal radius of the cavity, by acoustic measurements in pure argon, were presented in previous works [21,22].

Acoustic signal is generated inside the cavity by a capacitance-type acoustic transducer made ad-hoc for this technique powered by an ac signal from a wave synthesizer (3225B, HP) without bias voltage. Acoustic wave is detected at twice the driven frequency by other equal capacitance transducer externally polarized with 80 DCV with a Lock-In amplifier (SR850 DSP, SRS) operating at differential voltage mode and referenced to the wave generator. Phase sensitive detection performed by the Lock-In decomposes the received signal into two components, the in-phase "$u$" and the quadrature "$v$", which are interpolated to a Lorentzian shape function of frequency $f$ of the wave generator [23,24] as:

$$u + iv(f) = \frac{iAf}{(F^2 - f^2)} + B + C(f - f_{0n}) \qquad (1)$$

with $F = f_{0n} + ig_{0n}$, where $f_{0n}$ stands for resonance frequency and $g_{0n}$ stands for resonance halfwidth of the radial acoustic modes (0,$n$), and $A$, $B$, and $C$ are complex parameters.

Typical acoustic pressure levels of (1 to 20) mPa are generated, which produce signal amplitudes between (10 to 200) µV. More information about the design of acoustic transducers, acquisition instrumentation and procedures used for determination of the resonance acoustic signal were described elsewhere [25–27].

Temperature is determined as the mean of the readings of two 25.5 Ω standard platinum resistance thermometers (162D, Rosemount) plugged to an AC resistance bridge (F18, ASL), which are calibrated on the international temperature scale ITS-90 [28,29]. They are located in the northern and southern hemispheres of the acoustic cavity and the overall expanded ($k = 2$) uncertainty in temperature is better than 5 mK. Pressure is measured by two piezoelectric quartz transducers (Digiquartz 2003A-101, Paroscientific) for pressures from (0 to 2) MPa and (Digiquartz 43KR-101, Paroscientific) for pressures up to 20 MPa. They are calibrated against a dead-weight balance and the overall expanded ($k = 2$) in pressure is between (220 to 380) Pa for the explored range in this work.

Resonance frequencies $f_{0n}$ of the first five radial acoustic (0,$n$) modes are recorded from the highest measuring pressure and decreased in several steps down to nearly 0.1 MPa, obtaining about 8 to 11 points per isotherm for each mixture. The purely radial modes (0,$n$) are preferred due to their motion is normal to the wall of the resonance cavity, thus they are not affected by the viscous boundary layer and, in addition, they exhibit less sensitivity to imperfections of the geometry of the cavity.

The maximum measured pressure was set to 75 % of the saturation pressure $p_{sat}$ for those isotherms below the phase envelope to avoid the dispersion and absorption effects of precondensation on the speed of sound measurements [30]. This affects the isotherms $T$ = (273.16, 300, and 325) K for the mixture (0.60 $CO_2$ + 0.40 $C_3H_8$) with $p_{sat}$ = (1.17, 2.58 and 5.27) MPa, respectively, and isotherms $T$ = (273.16 and 300) K for the mixture (0.80 $CO_2$ + 0.20 $C_3H_8$) with $p_{sat}$ = (2.16 and 4.79) MPa, respectively. For the rest of temperatures, the amount of gas available in the sampling bottles limits the achievable pressure after several loads with the gas piston. On the other hand, the lowest pressure was chosen by the experience, as a compromise between a sufficiently low measuring state to get good accuracy when deriving the properties in the limit of zero pressure from the acoustic data, but with a resonance halfwidth not too wide that increases the fitting error of the resonance frequency to Equation (1).

Each measuring point consists of several repetitions of the acoustic resonance frequency $f_{0n}$ and halfwidth $g_{0n}$, the former are converted to their equivalent values at the reference temperatures $T$ from the experimental ones $T_{exp}$ using the estimated speed of sound $w(p,T)$ and $w(p,T_{exp})$ from REFPROP 10 [18] by:

$$f_{0n}(p,T) = f_{0n}(p,T_{exp}) \left( \frac{w(p,T)}{w(p,T_{exp})} \right) \tag{2}$$

and then, averaged to a single mean estimate of $f_{0n}$ and $g_{0n}$, to obtain the experimental speed of sound $w_{0n}(p,T)$ at the desired temperature $T$ and pressure $p$ for every mode and every mixture by:

$$w_{0n}(p,T) = \frac{2\pi a}{v_{0n}} (f_{0n} - \Delta f) \tag{3}$$

where $a = a(p,T)$ is the internal radius of the cavity and $v_{0n}$ is the zero of spherical Bessel first derivative of every $n^{th}$ mode. $\Delta f$ stands for the frequency correction, which is the sum of the frequency perturbations produced by different effects. Standard models were considered for the application of $\Delta f$ to our data sets, whose concrete expressions used in this work are specified elsewhere [31]. The overall frequency correction extends from −29 parts in $10^6$ at $p = 0.5$ MPa, $T = 325$ K, mode (0,6) for the (0.60 $CO_2$ + 0.40 $C_3H_8$) mixture until −107 parts in $10^6$ at $p = 0.12$ MPa, $T = 325$ K, mode (0,2) for the (0.80 $CO_2$ + 0.20 $C_3H_8$) mixture. The most significate contribution can be as high as −70 parts in $10^6$ from the thermal boundary layer perturbation, with a relatively small magnitude at the highest pressures of −40 parts in $10^6$ from the correction of the coupling of gas and shell motion due to the low value of the speed of sound for these mixtures, and non-depreciable terms from the ducts perturbation ranging from +10 to −25 parts in $10^6$.

Finally, perturbation on the acoustic resonance frequency and contribution to the halfwidth due to the relaxation phenomena was also assessed. It is already known that small rigid polyatomic molecules, such as carbon dioxide, exhibits long vibrational relaxations times $\tau$ that produces high absorption and dispersion of the sound wave [32–34]. On the contrary, more complex molecules such as propane have so small vibrational relaxation times $\tau$ that this effect is negligible [35–37]. At low densities, $\tau$ can be assumed as inversely proportional to the density, and for the case of mixtures of carbon dioxide (component 1) and propane (component 2), the vibrational relaxation time associated to like collisions of carbon dioxide molecules $\tau_{11}$ at $\rho = 1$ mol·m$^{-3}$ decrease from (280 to 180) μs in the temperature range of this research [32], while a more effective relaxation happens between unlike collisions of carbon dioxide and propane, with associated vibrational relaxation times $\tau_{12}$ about two orders of magnitude lower than $\tau_{11}$ [12,38].

Under the assumption that all carbon dioxide and propane molecules relax at unison following a parallel process each defined by a single relaxation time $\tau_k$, such that $\tau_k^{-1} = x_k/\tau_{kk} + x_j/\tau_{kj}$ is the vibrational relaxation time of molecule $k$ as a consequence of collisions between like molecules $\tau_{kk}$ and unlike molecules $\tau_{kj}$, the effect of the vibrational relaxation for carbon dioxide is clearly reduced due to the addition of propane. This is also reflected in the magnitude of the relative excess halfwidths, defined as $\Delta g_{0n}/f_{0n} = g_{\exp} - (g_{th} + g_0 + g_{bulk})/f_{0n}$, where $g_{th}$ stands for the contribution to the halfwidth of the thermal boundary layer, $g_0$ stands for the contribution to the halfwidth of the gas ducts, and $g_{bulk}$ stands for the viscothermal dissipation in the bulk of the fluid: the $\Delta g_{0n}/f_{0n}$ of pure carbon dioxide reported in the literature [32] is as high as 1200 parts in $10^6$ at the lowest experimental pressures reported of 0.4 MPa before allowance for vibrational relaxation, one order of magnitude larger than the 150 parts in $10^6$ for the higher mode (0,6) at the same state according to our values displayed in Figure 2.

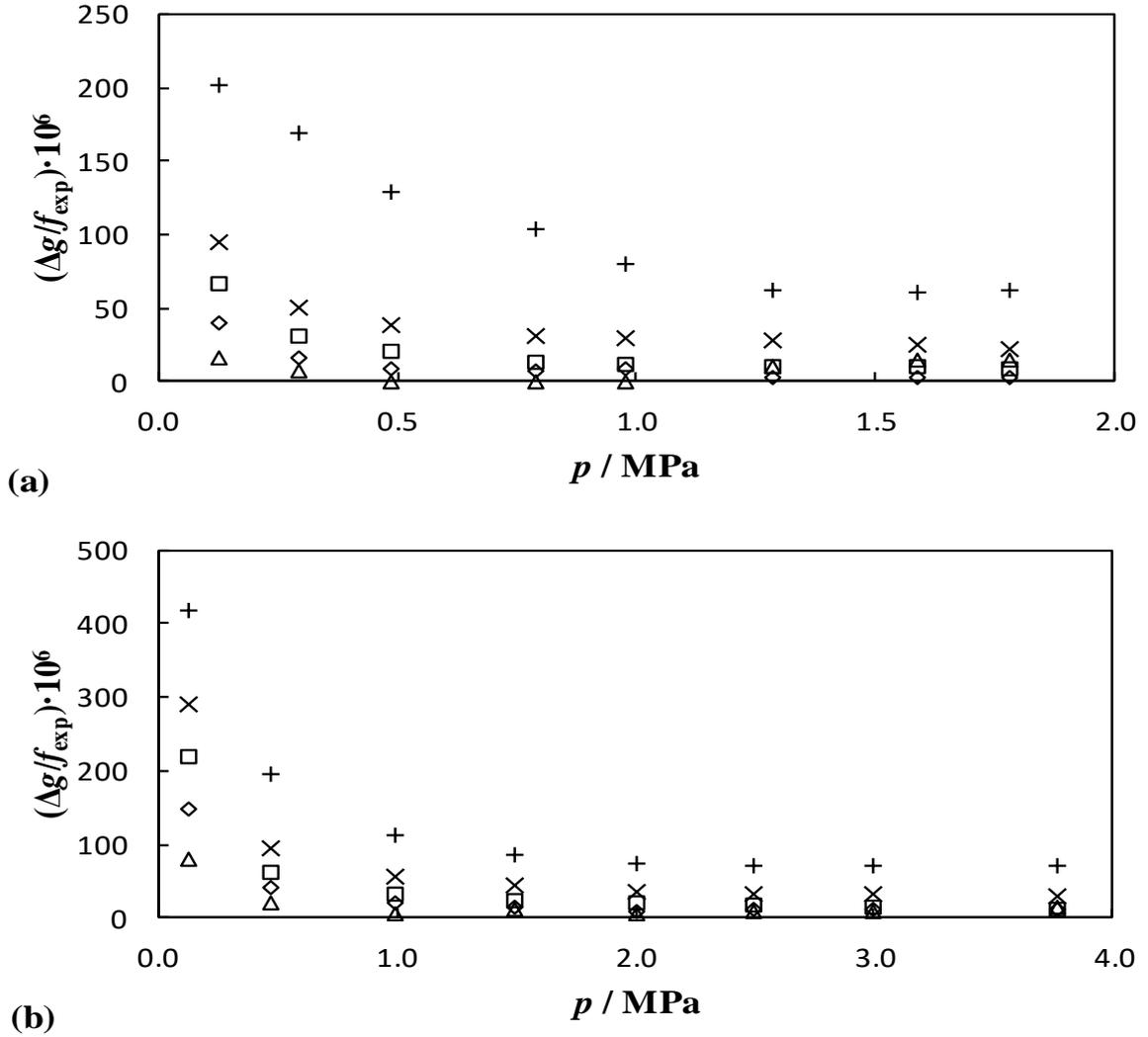

**Figure 2.** Relative excess halfwidths $(\Delta g/f)_{0n}$ for (a) the $(0.60\ CO_2 + 0.40\ C_3H_8)$ mixture and (b) for the $(0.80\ CO_2 + 0.20\ C_3H_8)$ mixture at $T = 375$ K and modes △ (0,2), ◇ (0,3), □ (0,4), × (0,5), + (0,6).

However, the excess halfwidth increasing trend as decreasing pressure indicates that some vibrational relaxation is taking place in our experiment. Following the procedure described by Estela-Uribe et al. [39], the contribution to the halfwidths $g_{vib}$ and the corresponding frequency perturbation $\Delta f_{vib}$ by vibrational relaxation on leading order are:

$$\frac{g_{vib}}{f} = \frac{\Delta g}{f} = \pi(\gamma-1)f\left[\Delta_1\tau_1 + \frac{\Delta_2\tau_2}{1+(2\pi f\tau_2)^2}\right] \quad (4)$$

$$\frac{\Delta f_{vib}}{f} = \frac{1}{2}(\gamma-1)\left[1-\frac{\Delta_1(1+3\gamma)}{4}\right]\Delta_1(2\pi f\tau_1)^2$$
$$+\frac{1}{2}(\gamma-1)\left[\frac{1}{1+(2\pi f\tau_2)^2}\right]\Delta_2(2\pi f\tau_2)^2 \quad (5)$$

where $\Delta_k = x_k C_{vib,k}/C_{p,m}$ is the vibrational contribution $C_{vib}$ to the molar isobaric heat capacity of the mixture $C_{p,m}$, with values of $\Delta_1 \geq 7\cdot10^{-2}$ and $\Delta_2 \geq 12\cdot10^{-2}$ and, therefore, with a significant contribution of the relaxation of both carbon dioxide and propane molecules at all the temperatures of this work. $C_{vib}$ was obtained from Plank-Einstein functions $C_{vib} = R\sum_i \frac{z_i^2 e^{z_i}}{(e^{z_i}-1)^2}$ with $z_i = (h_P\cdot v_i)/(k_B\cdot T)$ where $h_P$ stands for the Planck constant, $k_B$ stands for the Boltzmann constant, and $v_i$ are the $i$-th experimental vibrational frequency obtained from spectroscopy techniques for pure carbon dioxide [40] and propane [41]. Unfortunately, we have obtained really disparate results, with very small values of $\tau_{12}\cdot\rho \leq 10^{-10}$ s·(mol·m$^{-3}$) and very large values of $\tau_{21}\cdot\rho \geq 10$ s·(mol·m$^{-3}$), which suggest that the proposed model and assumptions are not suitable for the interpretation of the present data.

In any case, taken into account that the mode (0,6) is the only mode whose excess halfwidth $\Delta g_{06}$ is higher than $10^{-5}\cdot f_{06}$ along all the measuring pressure range and rise above the experimental expanded ($k = 2$) uncertainty of 250 parts in $10^6$ as decreasing the pressure, it was discarded for any following calculation. This is also supported by Figure 3, which shows that the speed of sound values, evaluated from mode (0,6), are not in agreement with the others at the same thermodynamic states. For the remaining modes, the relative dispersion of speed of sound data is lower than 40 parts in $10^6$, concluding that the applied acoustic model described above is correct.

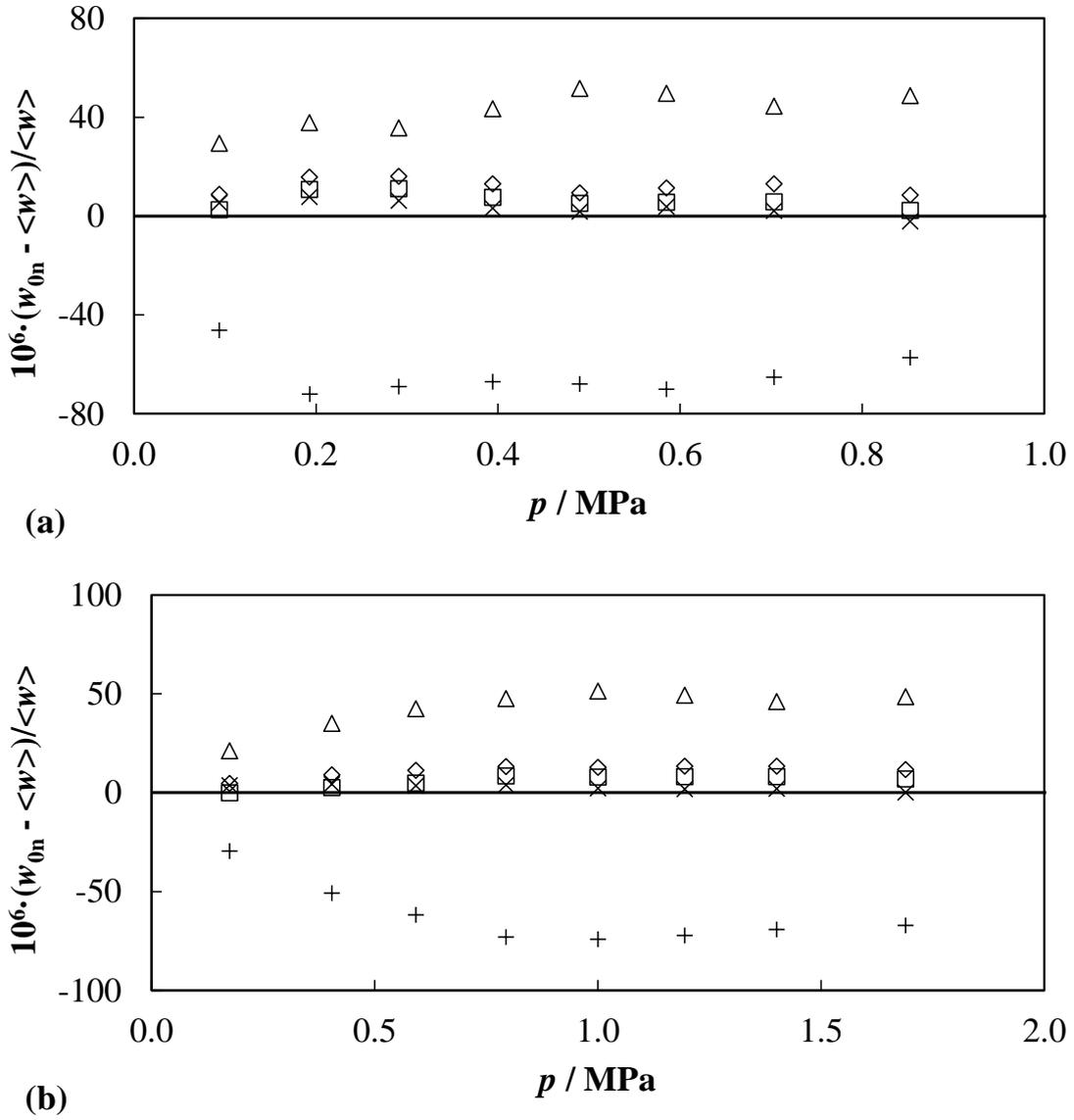

**Figure 3.** Relative dispersion of the speed of sound $\Delta w = (w_{0n} - \langle w \rangle)/\langle w \rangle$, where $\langle w \rangle$ is the mean value from modes (0,2) to (0,6), as a function of pressure for (a) the mixture (0.60 $CO_2$ + 0.40 $C_3H_8$) and (b) the mixture (0.80 $CO_2$ + 0.20 $C_3H_8$) at $T = 273.16$ K and modes △ (0,2), ◇ (0,3), □ (0,4), × (0,5), + (0,6)

A gross estimation of the extent of $\Delta f_{vib}$ may be made omitting the second term in the square bracket at the right side of Equation (4) and considering an effective vibrational relaxation time $\tau_{\text{eff}}$ for the mixture. Under this assumption, we obtained vibrational relaxation times at $\rho = 1$ mol·m$^{-3}$ of $\tau_{\text{eff}} = (1.38 \pm 0.13, 1.97 \pm 0.18, 1.69 \pm 0.15, 1.88 \pm 0.13$ and $1.79 \pm 0.13)$ μs for the temperatures $T = (273.16, 300, 325, 350$ and $375)$ K, respectively. Then, at the point of greatest value of $\Delta g_{0n}/f_{0n}$, measured for the mixture of higher carbon dioxide content (the relaxing gas) at the highest temperature and highest not neglected resonance mode, the mode (0,5): i) the relative excess halfwidth of mode (0,5) is reduced from (291 to 118) parts in $10^6$ after allowing for vibrational relaxation with the $\tau_{\text{eff}}$ reported, which would involve an increase of experimental expanded ($k = 2$) uncertainty from (250 to 276) parts in $10^6$ as much, and ii) the largest correction to the resonance frequency because of vibrational relaxation is always below 0.5 parts in $10^6$. Thus, we are confident that the vibrational relaxation is negligible in the present context.

**2.2.2 Cylindrical microwave resonator.**

Phase envelope is determined using a cylindrical microwave resonator based on the dielectric constant property change for the phase transition detection which is described in detail in [15], including the set-up of the equipment and the measurements of pure $CO_2$ in order to check the performance of the technique. A schematic view of the experimental technique is showed in the Fig. 4.

The measure cell consists of a cylindrical resonant cavity of 98.17 cm$^3$ of volume, with a sapphire tube, 15 cm long, located inside this cavity which contains the sample. The resonant modes of cylindrical cavity depend on the electrical properties of the sample. The cylindrical resonator is made of copper-zirconium (Luvata ZrK015) due to its low conductivity and mechanical simplicity. Two couplings are used to transfer microwave signal, using a vector network analyser (VNA N5230C, Agilent) configured to measure the complex scattering coefficient through the cavity and connected to the antennas.

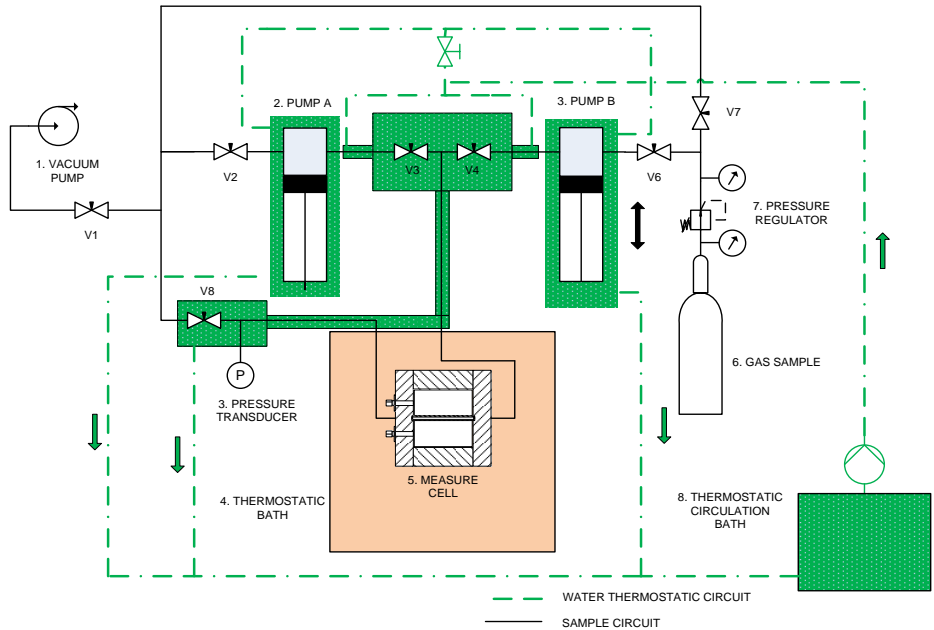

**Figure 4.** Schematic diagram of the technique for measuring phase envelope [15].

A precision thermostatic bath, which can work in a temperature range from −20°C to 150°C, is used to maintain the measure cell at isothermal conditions. Two syringe pumps A and B operate in a flow rate range from 1 μL/min to 90 mL/min at maximum pressures of 51.7 MPa. Additionally, an external circulation thermostatic bath controls the temperature of the syringe pumps and the sampling circuit. In order to homogenize the mixture, one of the syringe pumps works with positive displacement and another with negative displacement. A thermometer (F100, ASL) with two Pt-100 sensors was used for measuring temperature, and a pressure indicator (DPI-145, Druck) with an external transducer (PDCR 911-1756, Druck) was used for pressure measurements. Both devices were calibrated in our laboratory using equipment traceable to national standards. The corresponding calibration uncertainties are: $U(T) = 0.03$ K and $U(p) = (4.0 \cdot 10^{-4} \, (p/\mathrm{MPa}) + 0.001)$ MPa with a cover factor $k= 2$. To estimate the total uncertainty of temperature and pressure, when the bubble point or the dew point is measured, other contributions should be taken into account such as resolution, composition, repeatability and the polynomial adjustment for pressure, resulting in the values $U(T) = 0.04$ K and $U(p) =$ $(2.2 \cdot 10^{-3}\ (p/\text{MPa}) + 0.001)$ MPa with a cover factor $k = 2$.

Filling operation is done under homogenous phase conditions. During the isothermal experiment, liquid sample is maintained at constant temperature and the pressure is decreased until the first bubble occurs, and with the reverse procedure, dew point of a gas sample is measured. Phase transition presents a discontinuity in electric permittivity and therefore in resonant modes. The first transversal magnetic mode, $TM_{010}$, was chosen between the different resonant modes due to its high electric field strength in the axis of the cylindrical cavity, where sample is located, achieving an improvement in the sensitivity. Pressure of the sample is controlled by a thermostatic system in the surroundings of the resonant cylindrical cavity. High pressure equilibrium data of fluid mixtures can be measured with this technique which was previously checked measuring phase behavior of pure $CO_2$ and ($CO_2$ + $CH_4$) mixtures [15].

Accurate measurements of resonance frequency are performed by means of a Vector Network Analyser (VNA) from Agilent Technologies, with a frequency range from 300 kHz to 13.5 GHz and a maximum power from 3.98 to 12.59 mW. The VNA instrument measures complex transmission coefficient, $S_{21}$ (ratio of voltage transmitted to incident voltage), which is very important to fit resonance measured data. A model for an ideal resonator is the lumped-element series Resistor Inductor Capacitor (RLC) circuit that was used for the resonant mode $TM_{010}$. Equation 6 is used for modelling complex transmission coefficient as a function of frequency:

$$|S_{21}(f)| = \frac{|\overline{S_{21}}|}{\sqrt{1 + 4Q^2 \left(\frac{f}{f_0} - 1\right)^2}} \quad (6)$$

where $|S_{21}(f)|$ is the module of complex transmission coefficient which depends on frequency, $|\overline{S_{21}}|$ is the maximum value, $Q$ is the quality factor, $f$ is the frequency, and $f_0$ is the resonance frequency. For better fitting results, a variable change was implemented, and two sequential polynomic regressions are performed [42]. Microwave frequency measurements are performed with an uncertainty of 10 ppm which is negligible in comparison with the uncertainty of pressure or temperature measurements.

## 3. Results.

The 86 speed of sound data points were calculated from the average of the repeated acoustic resonance frequencies of the four experimental radial modes (0,2), (0,3), (0,4), and (0,5) using Equation (3), after rejection of mode (0,6) and application of the frequency perturbations of the described acoustic model, at temperatures $T$ = (273.16, 300, 325, 350, and 375) K and pressures $p$ up to 4 MPa for frequencies between (4.1 to 19.8) kHz. The values are reported in Table 2 for mixture (0.60 $CO_2$ + 0.40 $C_3H_8$) and Table 3 for mixture (0.80 $CO_2$ + 0.20 $C_3H_8$), together with the relative deviations from the AGA8-DC92 EoS [13,14] and GERG-2008 EoS [10,11] evaluated from REFPROP 10 [18].

**Table 2.** Gas-phase speed of sound $w_{exp}(p,T)$ for the mixture (0.60 $CO_2$ + 0.40 $C_3H_8$) with their relative expanded ($k$ = 2) uncertainty[(*)] and relative deviations $\Delta w_{EoS} = (w_{exp} - w_{EoS})/w_{EoS}$ from the speed of sound predicted by the AGA8-DC92 EoS [13,14] and GERG-2008 EoS [10,11]]. The exact mixture compositions with their uncertainties are provided in Table 1. Also, the experimental resonance frequencies $f_{0n}$ after corrections of the acoustic modes used to obtain $w_{exp}(p,T)$ are reported.

| $p$ / MPa | $w_{exp}$ / m·s$^{-1}$ | $10^6 \cdot \Delta w_{AGA}$ | $10^6 \cdot \Delta w_{GERG}$ | $f_{02}$ / Hz | $f_{03}$ / Hz | $f_{04}$ / Hz |
|---|---|---|---|---|---|---|
| | | | $T$ = 273.16 K | | | |
| 0.09341 | 247.590 | 327 | −116 | 4409.113 | 7580.203 | 10699.331 |
| 0.19282 | 245.847 | 672 | −150 | 4378.086 | 7526.790 | 10623.918 |
| 0.29079 | 244.094 | 1046 | −192 | 4346.866 | 7473.076 | 10548.113 |
| 0.39417 | 242.206 | 1501 | −238 | 4313.233 | 7415.230 | 10466.474 |
| 0.48965 | 240.424 | 1967 | −287 | 4281.500 | 7360.643 | 10389.425 |
| 0.58477 | 238.611 | 2500 | −336 | 4249.184 | 7305.106 | 10311.036 |
| 0.70301 | 236.298 | 3248 | −414 | 4207.966 | 7234.273 | 10211.056 |
| 0.85292 | 233.266 | 4396 | −522 | 4153.950 | 7141.377 | 10079.934 |
| | | | $T$ = 300.00 K | | | |
| 0.15300 | 257.606 | 346 | −330 | 4585.457 | 7883.318 | 11127.165 |
| 0.36130 | 254.726 | 955 | −475 | 4534.178 | 7795.127 | 11002.695 |
| 0.57462 | 251.696 | 1656 | −646 | 4480.217 | 7702.323 | 10871.693 |
| 0.76434 | 248.919 | 2323 | −850 | 4430.771 | 7617.307 | 10751.693 |
| 0.95467 | 246.076 | 3159 | −999 | 4380.153 | 7530.266 | 10628.836 |
| 1.18105 | 242.578 | 4248 | −1241 | 4317.879 | 7423.187 | 10477.683 |
| 1.38746 | 239.274 | 5387 | −1502 | 4259.020 | 7322.017 | 10334.872 |
| 1.58983 | 235.922 | 6695 | −1779 | 4199.338 | 7219.387 | 10190.022 |
| 1.76559 | 232.907 | 8002 | −2056 | 4145.647 | 7127.093 | 10059.761 |
| 1.97232 | 229.271 | 10018 | −2220 | 4080.919 | 7015.788 | 9902.633 |
| 2.16381 | 225.792 | 12393 | −2252 | 4018.964 | 6909.257 | 9752.341 |

*T* = 325.00 K

| | | | | | | |
|---|---|---|---|---|---|---|
| 0.09640 | 267.973 | 131 | −338 | 4767.777 | 8196.817 | 11569.687 |
| 0.28915 | 265.854 | 538 | −480 | 4730.077 | 8131.932 | 11478.082 |
| 0.49033 | 263.614 | 1021 | −626 | 4690.194 | 8063.324 | 11381.264 |
| 0.76321 | 260.516 | 1730 | −860 | 4635.028 | 7968.466 | 11247.383 |
| 0.98537 | 257.951 | 2398 | −1043 | 4589.362 | 7889.954 | 11136.543 |
| 1.19048 | 255.534 | 3042 | −1254 | 4546.324 | 7815.969 | 11032.082 |
| 1.37675 | 253.303 | 3681 | −1458 | 4506.602 | 7747.678 | 10935.689 |
| 1.58771 | 250.744 | 4511 | −1660 | 4461.030 | 7669.329 | 10825.099 |
| 1.79194 | 248.211 | 5351 | −1908 | 4415.925 | 7591.775 | 10715.655 |
| 2.01417 | 245.417 | 6422 | −2131 | 4366.201 | 7506.272 | 10594.968 |
| 2.22719 | 242.665 | 7477 | −2436 | 4317.191 | 7422.020 | 10476.048 |

*T* = 350.00 K

| | | | | | | |
|---|---|---|---|---|---|---|
| 0.13989 | 276.779 | 141 | −382 | 4922.710 | 8463.145 | 11945.596 |
| 0.27862 | 275.548 | 381 | −466 | 4900.804 | 8425.461 | 11892.395 |
| 0.48656 | 273.696 | 786 | −581 | 4867.823 | 8368.734 | 11812.337 |
| 0.79055 | 270.979 | 1497 | −703 | 4819.451 | 8285.557 | 11694.921 |
| 1.09466 | 268.220 | 2238 | −876 | 4770.334 | 8201.104 | 11575.696 |
| 1.39409 | 265.462 | 3000 | −1103 | 4721.239 | 8116.690 | 11456.535 |
| 1.68741 | 262.747 | 3878 | −1280 | 4672.908 | 8033.598 | 11339.272 |
| 2.04339 | 259.395 | 4971 | −1584 | 4613.226 | 7930.983 | 11194.444 |

*T* = 375.00 K

| | | | | | | |
|---|---|---|---|---|---|---|
| 0.12921 | 285.753 | −35 | −457 | 5080.289 | 8734.069 | 12328.004 |
| 0.29555 | 284.546 | 172 | −571 | 5058.806 | 8697.096 | 12275.806 |
| 0.48745 | 283.171 | 498 | −637 | 5034.331 | 8654.994 | 12216.400 |
| 0.79036 | 280.993 | 1037 | −760 | 4995.552 | 8588.309 | 12122.254 |

| p / MPa | $w_{exp}$ / m·s$^{-1}$ | $10^6 \cdot \Delta w_{AGA}$ | $10^6 \cdot \Delta w_{GERG}$ | $f_{02}$ / Hz | $f_{03}$ / Hz | $f_{04}$ / Hz |
|---|---|---|---|---|---|---|
| 0.97695 | 279.636 | 1346 | −887 | 4971.396 | 8546.794 | 12063.633 |
| 1.28414 | 277.391 | 1864 | −1133 | 4931.418 | 8478.047 | 11966.608 |
| 1.58734 | 275.211 | 2561 | −1247 | 4892.613 | 8411.324 | 11872.440 |
| 1.78169 | 273.795 | 2967 | −1389 | 4867.400 | 8367.962 | 11811.236 |

(*) Expanded uncertainties ($k = 2$): $U(p) = (2.2$ to $3.8) \cdot 10^{-4}$ MPa; $U(T) = 5$ mK; $U_r(w) = 2.5 \cdot 10^{-4}$ m·s$^{-1}$/ m·s$^{-1}$.

**Table 3.** Gas-phase speed of sound $w_{exp}(p,T)$ for the (0.80 $CO_2$ + 0.20 $C_3H_8$) mixture with their relative expanded ($k = 2$) uncertainty[*] and relative deviations $\Delta w_{EoS} = (w_{exp} - w_{EoS})/w_{EoS}$ from the speed of sound predicted by the AGA8-DC92 EoS [13,14] and GERG-2008 EoS [10,11]. The exact mixture compositions with their uncertainties are provided in Table 1. Also, the experimental resonance frequencies $f_{0n}$ after corrections of the acoustic modes used to obtain $w_{exp}(p,T)$ are reported.

| p / MPa | $w_{exp}$ / m·s$^{-1}$ | $10^6 \cdot \Delta w_{AGA}$ | $10^6 \cdot \Delta w_{GERG}$ | $f_{02}$ / Hz | $f_{03}$ / Hz | $f_{04}$ / Hz |
|---|---|---|---|---|---|---|
| | | | $T = 273.16$ K | | | |
| 0.17442 | 251.019 | 363 | 2 | 4470.188 | 7685.166 | 10847.475 |
| 0.40428 | 247.963 | 932 | 73 | 4415.718 | 7591.512 | 10715.295 |
| 0.59194 | 245.387 | 1514 | 130 | 4369.805 | 7512.599 | 10603.909 |
| 0.79421 | 242.523 | 2302 | 203 | 4318.791 | 7424.817 | 10480.006 |
| 1.00052 | 239.495 | 3301 | 279 | 4264.866 | 7332.021 | 10349.036 |
| 1.19464 | 236.536 | 4476 | 353 | 4212.127 | 7241.381 | 10221.087 |
| 1.40066 | 233.265 | 6042 | 441 | 4153.825 | 7141.200 | 10079.661 |
| 1.68967 | 228.411 | 9003 | 585 | 4067.358 | 6992.489 | 9869.768 |
| | | | $T = 300.00$ K | | | |
| 0.15849 | 262.372 | 297 | −67 | 4670.284 | 8029.171 | 11333.024 |
| 0.47740 | 259.135 | 915 | −46 | 4612.638 | 7929.994 | 11193.059 |
| 0.96014 | 254.060 | 2047 | −27 | 4522.271 | 7774.570 | 10973.679 |
| 1.46891 | 248.450 | 3546 | −28 | 4422.346 | 7602.783 | 10731.182 |
| 1.98784 | 242.410 | 5538 | −25 | 4314.769 | 7417.833 | 10470.149 |
| 2.51536 | 235.876 | 8241 | 2 | 4198.399 | 7217.803 | 10187.786 |
| 3.02555 | 229.073 | 11807 | 5 | 4077.225 | 7009.422 | 9893.877 |
| 3.23569 | 226.126 | 13791 | 105 | 4024.814 | 6919.284 | 9766.266 |
| | | | $T = 325.00$ K | | | |
| 0.11563 | 272.540 | 202 | −90 | 4849.031 | 8336.489 | 11766.834 |
| 0.48496 | 269.585 | 746 | −109 | 4796.427 | 8245.980 | 11639.070 |
| 0.99139 | 265.455 | 1622 | −146 | 4722.856 | 8119.471 | 11460.494 |
| 1.48068 | 261.370 | 2608 | −206 | 4650.086 | 7994.353 | 11283.901 |
| 1.98719 | 257.041 | 3796 | −278 | 4573.001 | 7861.788 | 11096.776 |
| 2.52348 | 252.338 | 5247 | −372 | 4489.244 | 7717.774 | 10893.490 |
| 2.98097 | 248.225 | 6659 | −464 | 4415.993 | 7591.821 | 10715.720 |
| 3.36390 | 244.709 | 7987 | −550 | 4353.399 | 7484.193 | 10563.793 |
| | | | $T = 350.00$ K | | | |
| 0.13074 | 281.726 | 35 | −251 | 5010.686 | 8614.397 | 12159.099 |
| 0.49094 | 279.431 | 455 | −278 | 4969.823 | 8544.082 | 12059.839 |
| 0.97289 | 276.333 | 1051 | −362 | 4914.653 | 8449.224 | 11925.936 |
| 1.47521 | 273.092 | 1794 | −435 | 4856.922 | 8349.947 | 11785.829 |
| 1.96912 | 269.877 | 2563 | −561 | 4799.653 | 8251.481 | 11646.845 |
| 2.51998 | 266.292 | 3577 | −656 | 4735.802 | 8141.689 | 11491.887 |
| 3.00725 | 263.099 | 4494 | −807 | 4678.924 | 8043.888 | 11353.840 |
| 3.36826 | 260.751 | 5282 | −859 | 4637.100 | 7971.996 | 11252.356 |
| | | | $T = 375.00$ K | | | |
| 0.12954 | 290.698 | −96 | −339 | 5168.194 | 8885.198 | 12541.332 |
| 0.47413 | 288.928 | 173 | −410 | 5136.681 | 8830.952 | 12464.752 |
| 0.99076 | 286.300 | 682 | −481 | 5089.874 | 8750.470 | 12351.131 |
| 1.49136 | 283.761 | 1203 | −592 | 5044.629 | 8672.669 | 12241.338 |
| 2.00136 | 281.202 | 1810 | −690 | 4999.026 | 8594.258 | 12130.672 |
| 2.49622 | 278.742 | 2442 | −801 | 4955.188 | 8518.882 | 12024.302 |
| 2.99485 | 276.307 | 3162 | −874 | 4911.809 | 8444.295 | 11919.023 |
| 3.76978 | 272.569 | 4215 | −1124 | 4845.198 | 8329.784 | 11757.369 |

(*) Expanded uncertainties ($k = 2$): $U(p) = (2.2 \text{ to } 3.8) \cdot 10^{-4}$ MPa; $U(T) = 5$ mK; $U_r(w) = 2.5 \cdot 10^{-4}$ m·s$^{-1}$/ m·s$^{-1}$.

The uncertainty budget for the speed of sound is listed in Table 4. The average of the expanded ($k = 2$) relative uncertainty of all the speed of sound datasets leads to the overall expanded ($k = 2$) uncertainty $U_r(w) = 250$ parts in $10^6$ (0.025 %). The procedure for the evaluation of each uncorrelated component of the uncertainty follows from the application of the Expression of Uncertainty in Measurement (GUM) [43,44] and was explained in other works [22,31]. The main contribution is due to the uncertainty of the speed of sound in argon, which amounts up to $u_r(\text{Ar}_{\text{EoS}}) = 100$ parts in $10^6$ (0.01 %), and has to be introduced in the dimensional characterization of the internal radius of the resonance cavity by acoustic measurements. The second contribution comes from the certified uncertainty of the gas composition, which results in a contribution of $u_r = 70$ parts in $10^6$ to all the measurements.

**Table 4.** Uncertainty budget for the speed of sound measurements $w_{exp}$. Unless otherwise specified, uncertainties are indicated with a coverage factor $k = 1$.

| Source | | Magnitude | Contribution to the speed of sound uncertainty, $10^6 \cdot u_r(w_{exp})$ |
|---|---|---|---|
| Temperature | Calibration | 0.002 K | |
| | Resolution | $7.2 \cdot 10^{-7}$ K | |
| | Repeatability | $3.5 \cdot 10^{-5}$ K | |
| | Gradient (across hemispheres) | $1.1 \cdot 10^{-3}$ K | |
| | Sum | 0.0025 K | 5.0 |
| Pressure | Calibration | $(3.75 \cdot 10^{-5} \cdot (p/\text{MPa}) + 1 \cdot 10^{-4})$ MPa | |
| | Resolution | $2.9 \cdot 10^{-5}$ MPa | |
| | Repeatability | $3.7 \cdot 10^{-6}$ MPa | |
| | Sum | (1.1 to 1.9)$\cdot 10^{-4}$ MPa | 6.4 |
| Gas composition | Purity | $1.4 \cdot 10^{-6}$ kg/mol | |
| | Molar mass | $6.0 \cdot 10^{-6}$ kg/mol | |
| | Sum | $6.2 \cdot 10^{-6}$ kg/mol | 73 |
| Radius from speed of sound in Ar | Temperature | $1.5 \cdot 10^{-9}$ m | |
| | Pressure | $1.6 \cdot 10^{-10}$ m | |
| | Gas Composition | $4.1 \cdot 10^{-9}$ m | |
| | Frequency fitting | $4.9 \cdot 10^{-7}$ m | |
| | Regression | $1.7 \cdot 10^{-6}$ m | |
| | Equation of State | $2.3 \cdot 10^{-6}$ m | |
| | Dispersion of modes | $2.9 \cdot 10^{-6}$ m | |
| | Sum | $4.2 \cdot 10^{-6}$ m | 98 |
| Frequency fitting | | 0.010 Hz | 1.7 |
| Dispersion of modes | | $3.0 \cdot 10^{-3}$ m·s$^{-1}$ | 12 |
| Sum of all contributions to $w_{exp}$ | | | 125 |
| $10^6 \cdot U_r(w_{exp})$ ($k = 2$) | | | 250 |

Concerning determination of the phase envelopes, the (0.60 $CO_2$ + 0.40 $C_3H_8$) mixture was measured in the temperature range from 229.15 K to 329.60 K and in the pressure range from 0.2 MPa to 6.8 MPa, including the retrograde zone considering the critical reference point calculated with the GERG 2008 [10,11], using the software REFPROP [18] whereas the range of measurements for the (0.80 $CO_2$ + 0.20 $C_3H_8$) mixture was from 233.15 K to 310.36 K in temperature, and from 0.4 MPa to 6.7 MPa in pressure. Table 5 and Table 6 contain the experimental data and their comparison with the calculated pressure using GERG-2008 EoS. In addition, pressure and temperature uncertainties for these data were calculated taking into account the calibration and the repeatability of the measurements, obtaining the values given at the bottom of the tables.

**Table 5.** Phase equilibria data[a] for the (0.60 $CO_2$ + 0.40 $C_3H_8$) mixture and comparison with data calculated using the GERG-2008 EoS [11]. The exact mixture compositions with their uncertainties are provided in Table 1.

| $T_{exp}$ / K | Set-up | $p_{exp}$ / MPa | $p_{GERG}$ / MPa | $\Delta p$ / %[b] | Set-up | $p_{exp}$ / MPa | $p_{GERG}$ / MPa | $\Delta p$ / %[b] |
|---|---|---|---|---|---|---|---|---|
| 229.37 | Bubble | 0.738 | 0.738 | −0.10 | Dew | 0.227 | 0.226 | −0.08 |
| 233.35 | Bubble | 0.845 | 0.848 | 0.36 | Dew | 0.272 | 0.27 | −0.67 |
| 238.36 | Bubble | 1.005 | 1.002 | −0.33 | Dew | 0.334 | 0.334 | 0.02 |
| 243.35 | Bubble | 1.180 | 1.175 | −0.46 | Dew | 0.411 | 0.409 | −0.58 |
| 248.34 | Bubble | 1.358 | 1.367 | 0.68 | Dew | 0.496 | 0.496 | −0.07 |
| 253.35 | Bubble | 1.594 | 1.581 | −0.85 | Dew | 0.595 | 0.598 | 0.60 |
| 258.34 | Bubble | 1.825 | 1.815 | −0.56 | Dew | 0.717 | 0.715 | −0.19 |
| 263.34 | Bubble | 2.070 | 2.073 | 0.13 | Dew | 0.846 | 0.85 | 0.54 |
| 268.34 | Bubble | 2.337 | 2.353 | 0.67 | Dew | 1.004 | 1.005 | 0.05 |
| 273.34 | Bubble | 2.64 | 2.657 | 0.62 | Dew | 1.176 | 1.181 | 0.44 |
| 278.33 | Bubble | 2.982 | 2.984 | 0.07 | Dew | 1.388 | 1.381 | −0.53 |
| 283.38 | Bubble | 3.342 | 3.339 | −0.08 | Dew | 1.599 | 1.61 | 0.68 |
| 288.4 | Bubble | 3.711 | 3.716 | 0.15 | Dew | 1.87 | 1.868 | −0.13 |
| 293.41 | Bubble | 4.128 | 4.116 | −0.28 | Dew | 2.158 | 2.159 | 0.04 |
| 298.22 | Bubble | 4.539 | 4.521 | −0.4 | Dew | 2.476 | 2.475 | −0.05 |
| 303.36 | Bubble | 4.966 | 4.972 | 0.12 | Dew | 2.867 | 2.858 | −0.33 |
| 308.46 | Bubble | 5.439 | 5.435 | −0.09 | Dew | 3.294 | 3.292 | −0.05 |
| 313.47 | Bubble | 5.915 | 5.893 | −0.37 | Dew | 3.783 | 3.784 | 0.02 |
| 318.49 | Bubble | 6.329 | 6.338 | 0.14 | Dew | 4.359 | 4.362 | 0.06 |
| 323.49 | Bubble | 6.721 | 6.719 | −0.03 | Dew | 5.068 | 5.063 | −0.10 |
| 324.49 | Bubble | 6.773 | 6.778 | 0.08 | Dew | 5.223 | 5.227 | 0.08 |
| 325.49 | Bubble | 6.822 | 6.827 | 0.07 | Dew | 5.394 | 5.402 | 0.14 |
| 326.47 | Bubble | 6.859 | 6.861 | 0.03 | Dew | 5.579 | 5.589 | 0.18 |
| 327.44 | Bubble | 6.874 | 6.873 | 0.01 | Dew | 5.807 | 5.794 | −0.23 |
| 328.41 | Dewc | 6.857 | 6.861 | 0.06 | Dew | 6.043 | 6.034 | −0.14 |
| 328.53 | Dewc | 6.865 | 6.856 | −0.14 | Dew | 6.056 | 6.068 | 0.19 |
| 328.73 | Dewc | 6.844 | 6.844 | 0 | Dew | 6.111 | 6.127 | 0.26 |
| 328.93 | Dewc | 6.834 | 6.827 | −0.1 | Dew | 6.201 | 6.192 | −0.14 |
| 329.53 | Dewc | 6.724 | 6.702 | −0.32 | Dew | 6.459 | 6.459 | 0.01 |
| 329.60 | Dewc | 6.681 | 6.655 | −0.39 | Dew | 6.538 | 6.524 | −0.22 |

[a] Expanded uncertainty ($k = 2$): $U(T) = 0.04$ K; $U(p) = (2.2 \cdot 10^{-3} \, (p/\text{MPa}) + 0.001)$ MPa.

[b] $\delta p / \% = 100 \cdot (p_{\text{exp}}/p_{\text{calc}} - 1)$.

[c] Retrograde dew point

**Table 6.** Phase equilibria data[a] for the (0.80 $CO_2$ + 0.20 $C_3H_8$) mixture and comparison with data calculated using the GERG-2008 EoS [11]. The exact mixture compositions with their uncertainties are provided in Table 1.

| $T_{exp}$ / K | Set-up | $p_{exp}$ / MPa | $p_{GERG}$ / MPa | $\Delta p$ / %[b] | Set-up | $p_{exp}$ / MPa | $p_{GERG}$ / MPa | $\Delta p$ / %[b] |
|---|---|---|---|---|---|---|---|---|
| 233.36 | Bubble | 0.923 | 0.921 | 0.17 | Dew | 0.497 | 0.495 | 0.4 |
| 238.37 | Bubble | 1.095 | 1.098 | −0.27 | Dew | 0.609 | 0.612 | −0.49 |
| 243.36 | Bubble | 1.296 | 1.297 | −0.08 | Dew | 0.748 | 0.748 | 0 |
| 248.35 | Bubble | 1.522 | 1.521 | 0.07 | Dew | 0.903 | 0.908 | −0.55 |
| 253.36 | Bubble | 1.775 | 1.772 | 0.16 | Dew | 1.09 | 1.094 | −0.37 |
| 258.35 | Bubble | 2.056 | 2.05 | 0.28 | Dew | 1.305 | 1.308 | −0.23 |
| 263.35 | Bubble | 2.362 | 2.359 | 0.15 | Dew | 1.545 | 1.555 | −0.67 |
| 268.34 | Bubble | 2.707 | 2.697 | 0.36 | Dew | 1.839 | 1.838 | 0.05 |
| 273.32 | Bubble | 3.072 | 3.068 | 0.14 | Dew | 2.162 | 2.159 | 0.14 |
| 278.34 | Bubble | 3.478 | 3.475 | 0.08 | Dew | 2.526 | 2.527 | −0.04 |
| 283.33 | Bubble | 3.911 | 3.916 | −0.12 | Dew | 2.954 | 2.942 | 0.41 |
| 288.33 | Bubble | 4.409 | 4.393 | 0.36 | Dew | 3.411 | 3.412 | −0.03 |
| 293.34 | Bubble | 4.921 | 4.908 | 0.26 | Dew | 3.916 | 3.945 | −0.74 |
| 298.33 | Bubble | 5.471 | 5.456 | 0.27 | Dew | 4.544 | 4.547 | −0.07 |
| 303.34 | Bubble | 6.038 | 6.035 | 0.05 | Dew | 5.265 | 5.241 | 0.46 |
| 308.34 | Bubble | 6.614 | 6.602 | 0.19 | Dew | 6.073 | 6.074 | −0.02 |
| 308.54 | Bubble | 6.632 | 6.622 | 0.15 | Dew | 6.117 | 6.113 | 0.07 |
| 308.95 | Bubble | 6.684 | 6.662 | 0.33 | Dew | 6.195 | 6.194 | 0.01 |
| 309.16 | Bubble | 6.695 | 6.681 | 0.21 | Dew | 6.244 | 6.237 | 0.11 |
| 309.35 | Bubble | 6.714 | 6.697 | 0.25 | Dew | 6.285 | 6.277 | 0.12 |
| 309.55 | Bubble | 6.726 | 6.713 | 0.19 | Dew | 6.328 | 6.32 | 0.12 |
| 309.96 | Bubble | 6.746 | 6.741 | 0.08 | Dew | 6.42 | 6.414 | 0.09 |
| 310.16 | Bubble | 6.747 | 6.751 | −0.06 | Dew | 6.451 | 6.463 | −0.19 |
| 310.36 | Bubble | 6.772 | 6.759 | 0.19 | Dew | 6.51 | 6.516 | −0.09 |

[a] Expanded uncertainty ($k = 2$): $U(T) = 0.04$ K; $U(p) = (2.2 \cdot 10^{-3} \, (p/\text{MPa}) + 0.001)$ MPa.

[b] $\delta p$ / % = $100 \cdot (p_{exp}/p_{calc} - 1)$.

**4. Discussion.**

**4.1 Speed of sound.**

Relative deviations of the experimental speed of sound data from the calculated values by AGA8-DC92 EoS [13,14] and GERG-2008 EoS [10,11] are listed in Table 2 and Table 3 (in the former section), and also depicted in Figure 5 and Figure 6.

Nearly all the differences, when comparing to the GERG-2008 EoS [10,11], remain outside $U_r(w_{exp}) = 250$ parts in $10^6$ (0.025 %), with the exceptions of the deviations at the isotherm of 300 K along all pressures and the isotherms of 273.16 K and 325 K up to intermediate pressures for the (0.80 $CO_2$ + 0.20 $C_3H_8$) mixture, and the state points limited to the lower pressure range for all the temperatures and both mixtures, which approach to zero while decreasing pressure. This good agreement with the values close to zero pressure is a good indicator of the stability of the mixture composition during the measuring process, so we conclude that the measurements are not affected by the absorption and desorption of the gas with the cavity wall material, condensation during the gas filling procedure or stratification of the mixtures within the sampling gas bottles.

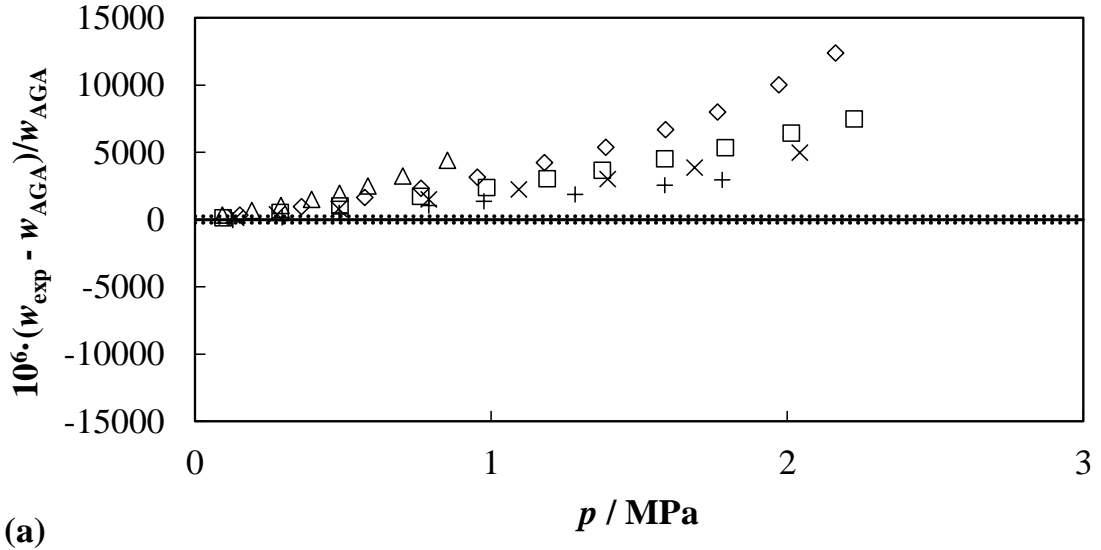

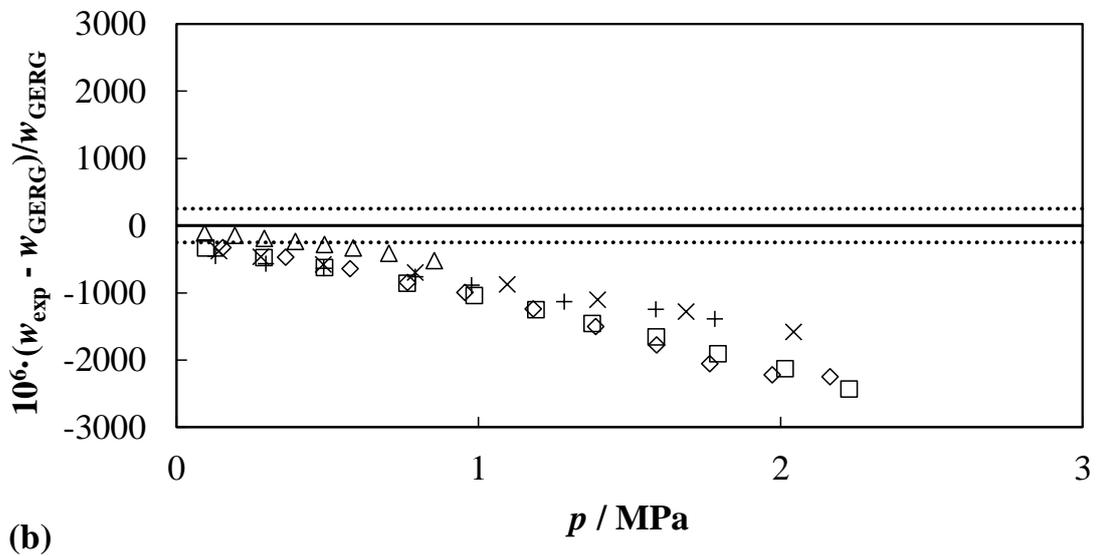

**Figure 5.** Relative deviations of experimental speed of sound from speed of sound values calculated from (a) AGA8-DC92 EoS [13,14] and (b) GERG-2008 EoS [10,11] as function of pressure for the (0.60 $CO_2$ + 0.40 $C_3H_8$) mixture at temperatures $T$ = △ 273.16 K, ◇ 300 K, □ 325 K, × 350 K, + 375 K. Dotted line depicts the expanded ($k$ = 2) uncertainty of the experimental speed of sound.

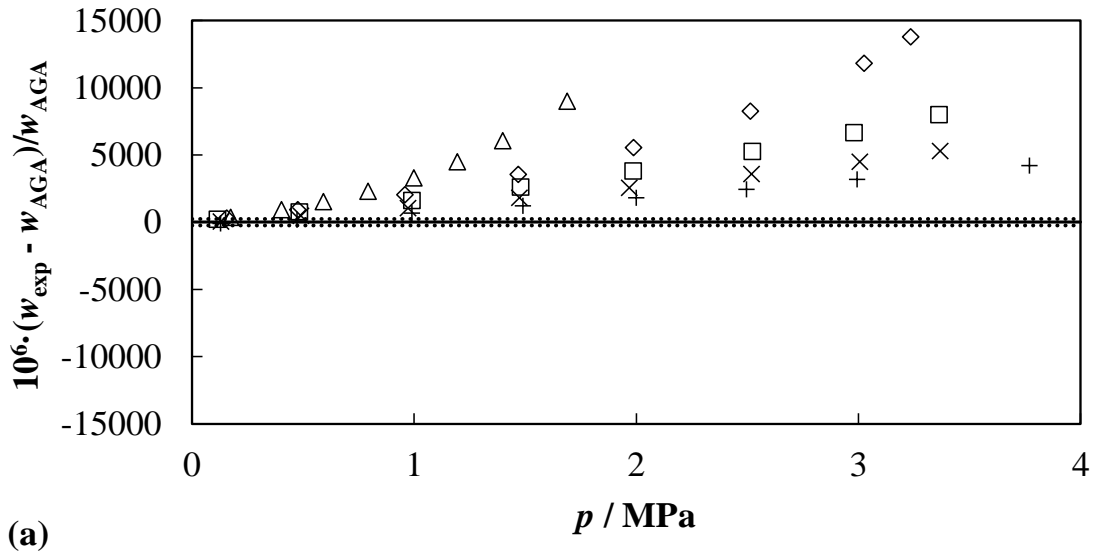

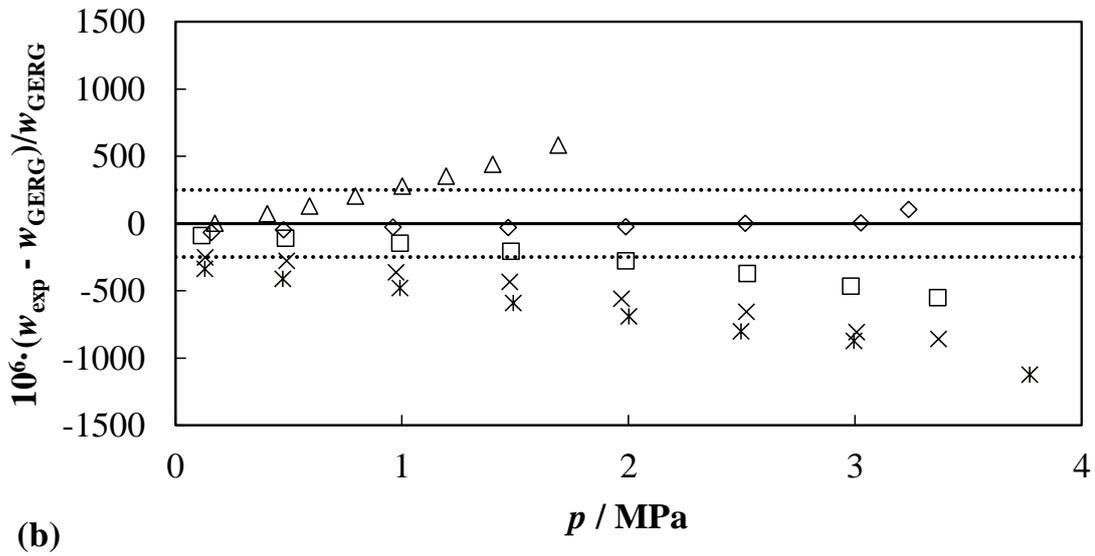

**Figure 6.** Relative deviations of experimental speed of sound from speed of sound values calculated from (a) AGA8-DC92 EoS [13,14] and (b) GERG-2008 EoS [10,11] as function of pressure for the (0.80 $CO_2$ + 0.20 $C_3H_8$) mixture at temperatures $T$ = △ 273.16 K, ◇ 300 K, □ 325 K, × 350 K, + 375 K. Dotted line depicts the expanded ($k$ = 2) uncertainty of the experimental speed of sound.

However, discrepancies are well explained within the declared expanded ($k = 2$) uncertainty of the GERG-2008 EoS [10,11] in speed of sound $U_r(w_{GERG}) = 10000$ parts in $10^6$ (1.0 %) for binary mixtures with no departure function and only adjusted reducing function. Though the AGA8-DC92 EoS [13,14] does not report any uncertainty for this kind of mixtures, even assuming the same model uncertainty than the GERG-2008 EoS [10,11], deviations are also within these limits.

Deviations increase with increasing pressure, with systematic negative trends with respect to GERG-2008 EoS [10,11] that decrease with decreasing $x_{C3H8}$, in contrast to positive differences according to AGA8-DC92 EoS [13,14] that increase with decreasing $x_{C3H8}$. Average absolute deviations $\Delta_{AAD} = \left[ N^{-1} \sum_{i=1}^{N} \left| (w_{exp} - w_{EoS})/w_{EoS} \right|_i \right]$ between experimental and calculated values are better than 0.093 % for the (0.60 $CO_2$ + 0.40 $C_3H_8$) mixture and less than 0.035 % for the (0.80 $CO_2$ + 0.20 $C_3H_8$) mixture with respect to the GERG-2008 EoS [10,11]. The corresponding $\Delta_{AAD}$ using AGA8-DC92 EoS [13,14] are (0.27 and 0.34) %, respectively. These results are consistent with the declared uncertainty of the two equations.

Each squared speed of sound data set $w^2(p_i,T)$ at several pressures $p_i$ and referenced to the same temperature $T$ were analyzed by fitting to a polynomial function of the pressure:

$$w^2(p,T) = A_0(T) + A_1(T)p + A_2(T)p^2 + A_3(T)p^3 \qquad (7)$$

with the adjustable parameters $A_i$ listed in Table 7 together with their expanded ($k = 2$) uncertainty $U(A_i)$ computed by the Monte Carlo method [44].

**Table 7.** Fitting parameters $A_i(T)$ of Eq. (7) with their corresponding expanded ($k = 2$) uncertainties, determined by the Monte Carlo method, and the root mean square ($\Delta_{RMS}$) of the residuals of the fitting.

| $T$ / K | $A_0(T)$ / $m^2 \cdot s^{-2}$ | $10^5 \cdot A_1(T)$ / $m^2 \cdot s^{-2} \cdot Pa^{-1}$ | $10^{12} \cdot A_2(T)$ / $m^2 \cdot s^{-2} \cdot Pa^{-2}$ | $10^{18} \cdot A_3(T)$ / $m^2 \cdot s^{-2} \cdot Pa^{-3}$ | $\Delta_{RMS}$ of residuals / ppm |
|---|---|---|---|---|---|
| (0.60 $CO_2$ + 0.40 $C_3H_8$) | | | | | |
| 273.16 | 62092 ± 13 | −842.9 ± 6.4 | −671 ± 65 | | 18 |
| 300.00 | 67426 ± 16 | −693.2 ± 5.6 | −258 ± 54 | −24 ± 15 | 40 |
| 325.00 | 72368 ± 10 | −580.9 ± 2.0 | −108.8 ± 8.1 | | 20 |
| 350.00 | 77287 ± 15 | −487.4 ± 3.3 | −10 ± 15 | | 19 |
| 375.00 | 82186 ± 18 | −412.4 ± 4.5 | 39 ± 23 | | 35 |
| (0.80 $CO_2$ + 0.20 $C_3H_8$) | | | | | |
| 273.16 | 64152 ± 17 | −650.2 ± 7.4 | −200 ± 89 | −88 ± 31 | 7 |
| 300.00 | 69675 ± 14 | −525.2 ± 3.7 | −73 ± 25 | −23.2 ± 4.8 | 25 |
| 325.00 | 74777 ± 11 | −431.5 ± 1.4 | −33.4 ± 3.9 | | 3 |
| 350.00 | 79846 ± 12 | −361.4 ± 1.6 | 27.4 ± 4.4 | | 37 |
| 375.00 | 84899 ± 13 | −301.8 ± 1.6 | 54.3 ± 3.9 | | 27 |

A second order polynomial was required for most of the isotherms, except for $T = 273.16$ K for the mixture (0.60 $CO_2$ + 0.40 $C_3H_8$) and $T = (273.16$ and $300)$ K for the mixture (0.80 $CO_2$ + 0.20 $C_3H_8$), in order to meet the criteria of residuals within the experimental uncertainty, random distribution of the residuals when represented versus the pressure (independent variable) and versus the experimental speed of sound (dependent variable), and statistical significance of the regression parameters as given by the $p$-test. The estimated uncertainties of each speed of sound value are used as weights in this regression. Relative residuals of the fitting are plotted in Figure 7 for the two mixtures, with a root mean square $\Delta_{RMS}$ less than 30 parts in $10^6$ in all the cases.

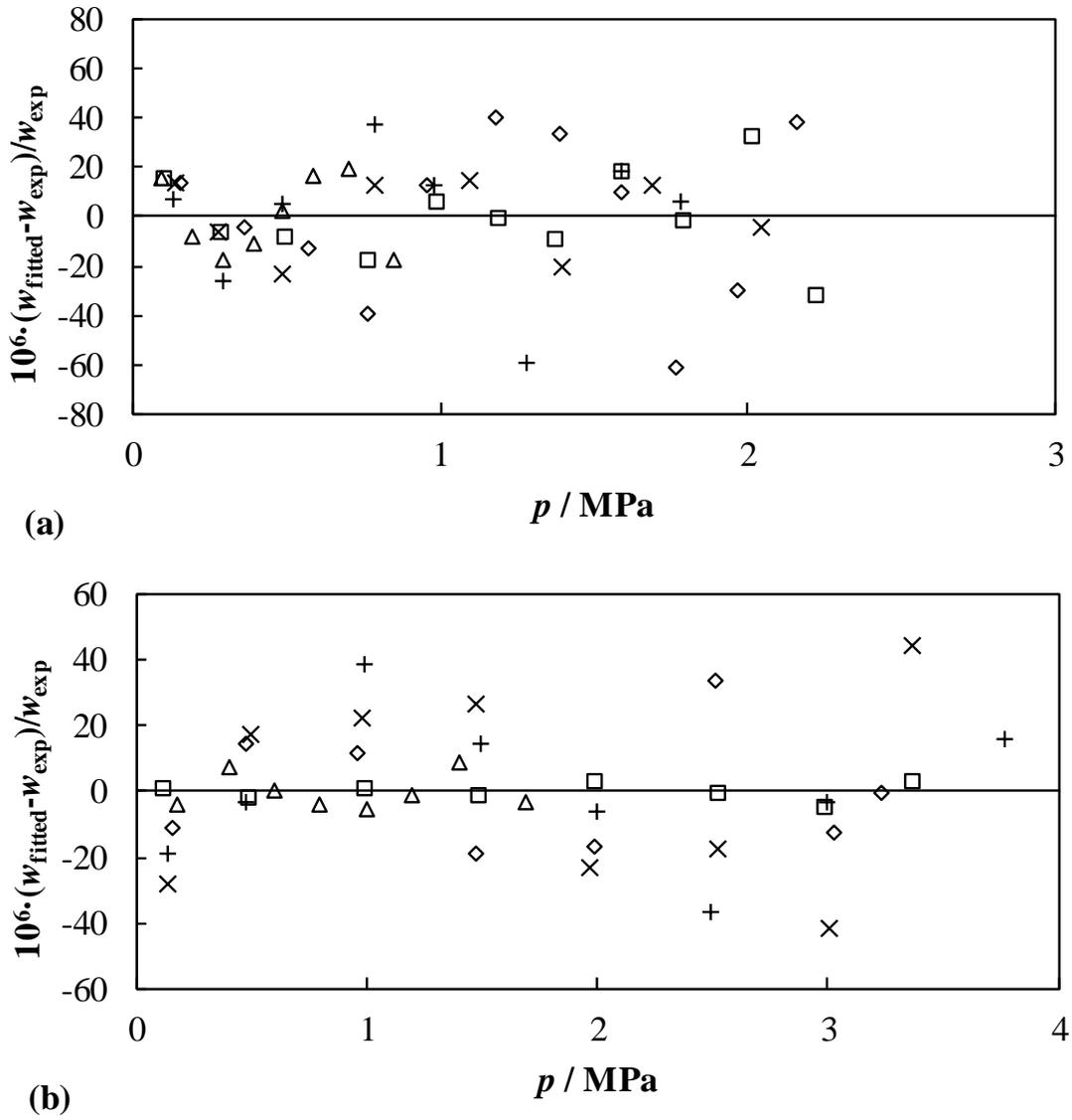

**Figure 7.** Speed of sound residuals from Eq. (7) as a function of the pressure for (a) the (0.60 $CO_2$ + 0.40 $C_3H_8$) mixture and (b) the (0.80 $CO_2$ + 0.20 $C_3H_8$) mixture at temperatures $T =$ △ 273.16 K, ◇ 300 K, □ 325 K, × 350 K, + 375 K.

### 4.2 Perfect-gas heat capacities.

Molar perfect-gas heat capacities at constant pressure $C_{\text{p,m}}^{\text{pg}}$ are obtained from the fitting parameters $A_0(T)$ of Table 7 by:

$$\gamma^{\text{pg}} = \frac{C_{\text{p,m}}^{\text{pg}}}{C_{\text{V,m}}^{\text{pg}}} = \frac{C_{\text{p,m}}^{\text{pg}}}{C_{\text{p,m}}^{\text{pg}} - R} = \frac{M}{RT} A_0 \qquad (8)$$

where $\gamma^{\text{pg}}$ is the heat capacity ratio as perfect-gas "$pg$" (zero pressure), $M$ is the molar mass of the mixture (Table 1) and $R$ is the exact value of the molar gas constant [45]. Table 8 reports the experimental results $\gamma^{\text{pg}}$ and $C_{\text{p,m}}^{\text{pg}}$ together with their expanded ($k = 2$) uncertainty $U(\gamma^{\text{pg}})$ and $U(C_{\text{p,m}}^{\text{pg}})$, which is better than 0.31 % for the (0.60 $CO_2$ + 0.40 $C_3H_8$) mixture and less than 0.17 % for the (0.80 $CO_2$ + 0.20 $C_3H_8$) mixture, and the comparison with the two reference EoS. This uncertainty includes the contributions of temperature, molar mass, and fitted $A_0$.

**Table 8.** Adiabatic coefficient $\gamma^{\text{pg}}$, isobaric heat capacity $C_{\text{p,m}}^{\text{pg}}$, acoustic second virial coefficient $\beta_a$, and acoustic third virial coefficient $\gamma_a$ for the ($CO_2$ + $C_3H_8$) mixtures determined in this work, with their corresponding relative expanded ($k = 2$) uncertainty, and comparison with AGA8-DC92 and GERG-2008 EoS. The superscript $pg$ indicates perfect-gas property.

| $T$ / K | $\gamma^{\text{pg}}$ | $10^2 \cdot U_r(\gamma^{\text{pg}})$ | $10^2 \cdot \Delta\gamma^{\text{pg}}_{\text{AGA}}{}^{(*)}$ | $10^2 \cdot \Delta\gamma^{\text{pg}}_{\text{GERG}}{}^{(*)}$ | $C_{\text{p,m}}^{\text{pg}}$ / J·mol$^{-1}$·K$^{-1}$ | $10^2 \cdot U_r(C_{\text{p,m}}^{\text{pg}})$ | $10^2 \cdot \Delta C_{\text{p,m}}^{\text{pg}}{}_{\text{AGA}}{}^{(*)}$ | $10^2 \cdot \Delta C_{\text{p,m}}^{\text{pg}}{}_{\text{GERG}}{}^{(*)}$ |
|---|---|---|---|---|---|---|---|---|
| (0.60 $CO_2$ + 0.40 $C_3H_8$) | | | | | | | | |
| 273.16 | 1.20412 | 0.042 | −0.0066 | −0.029 | 49.047 | 0.25 | 0.032 | 0.14 |
| 300.00 | 1.19058 | 0.044 | −0.021 | −0.055 | 51.943 | 0.27 | 0.11 | 0.29 |
| 325.00 | 1.17955 | 0.039 | −0.019 | −0.060 | 54.621 | 0.26 | 0.11 | 0.33 |
| 350.00 | 1.16974 | 0.041 | −0.025 | −0.066 | 57.298 | 0.29 | 0.15 | 0.39 |
| 375.00 | 1.16097 | 0.043 | −0.043 | −0.080 | 59.968 | 0.31 | 0.27 | 0.49 |

| | | | (0.80 $CO_2$ + 0.20 $C_3H_8$) | | | | | |
|---|---|---|---|---|---|---|---|---|
| 273.16 | 1.24359 | 0.034 | 0.011 | −0.0012 | 42.448 | 0.17 | −0.047 | 0.0042 |
| 300.00 | 1.22982 | 0.028 | 0.014 | −0.0068 | 44.493 | 0.15 | −0.057 | 0.029 |
| 325.00 | 1.21834 | 0.025 | 0.0082 | −0.018 | 46.394 | 0.14 | −0.035 | 0.081 |
| 350.00 | 1.20801 | 0.025 | −0.010 | −0.038 | 48.286 | 0.15 | 0.052 | 0.18 |
| 375.00 | 1.19882 | 0.025 | −0.032 | −0.056 | 50.134 | 0.15 | 0.16 | 0.28 |

| | $10^7 \cdot \beta_a$ / $m^3 \cdot mol^{-1}$ | $10^2 \cdot U_r(\beta_a)$ | $10^2 \cdot \Delta\beta_{a,AGA}^{(*)}$ | $10^2 \cdot \Delta\beta_{a,GERG}^{(*)}$ | $10^{10} \cdot \gamma_a$ / $(m^3 \cdot mol^{-1})^2$ | $10^2 \cdot U_r(\gamma_a)$ | $10^2 \cdot \Delta\gamma_{a,AGA}^{(*)}$ | $10^2 \cdot \Delta\gamma_{a,GERG}^{(*)}$ |
|---|---|---|---|---|---|---|---|---|
| | | | (0.60 $CO_2$ + 0.40 $C_3H_8$) | | | | | |
| 273.16 | −3083 | 0.78 | −5.0 | −0.44 | 155 | 35 | −40 | −51 |
| 300.00 | −2564 | 0.78 | −5.1 | 0.74 | 243 | 21 | −12 | −15 |
| 325.00 | −2169.2 | 0.34 | −5.0 | 1.3 | 229.7 | 3.6 | −7.1 | −11 |
| 350.00 | −1835.1 | 0.69 | −5.2 | 1.4 | 231.4 | 7.1 | 5.4 | −2.1 |
| 375.00 | −1564.6 | 1.1 | −4.9 | 2.0 | 222 | 12 | 13 | 2.6 |
| | | | (0.80 $CO_2$ + 0.20 $C_3H_8$) | | | | | |
| 273.16 | −2302 | 1.1 | −2.9 | −0.019 | 263 | 27 | 79 | 28 |
| 300.00 | −1880.3 | 0.70 | −3.7 | 0.36 | 215 | 10 | 26 | 15 |
| 325.00 | −1559.4 | 0.33 | −4.4 | −0.0071 | 160.7 | 2.4 | 3.2 | −6.0 |
| 350.00 | −1317.1 | 0.46 | −3.7 | 1.0 | 166.3 | 2.8 | 19 | 6.2 |
| 375.00 | −1108.5 | 0.52 | −3.2 | 1.7 | 159.8 | 2.9 | 25 | 12 |

(*) $\Delta X_{EoS} = (X_{exp} - X_{EoS})/X_{EoS}$ with $X = \gamma^{pg}$, $C_{p,m}^{pg}$, $\beta_a$, $\gamma_a$; and EoS = AGA8-DC92 [13,14], GERG-2008 [10,11].

Perfect-gas heat capacities for the ($CO_2$ + $C_3H_8$) mixture evaluated from the models are obtained from the spectroscopy data of the unpublished work of Chao et al. [46] for pure carbon dioxide with an expanded ($k = 2$) uncertainty of 0.02 %, and the perfect-gas heat capacity from the spectroscopy data of Chao et al. [47] for pure propane with an expanded ($k = 2$) uncertainty between (0.3 to 0.5) %, which slightly differ from the speed of sound data of Trusler et al. [35] used in the development of the current reference EoS for pure propane [20]. Thus, the estimated expanded ($k = 2$) uncertainty of $C_{p,m}^{pg}$ from the two models is by about 0.2 % for the (0.60 $CO_2$ + 0.40 $C_3H_8$) mixture and 0.1 % for the (0.80 $CO_2$ + 0.20 $C_3H_8$) mixture. As shown in Table 8, the relative deviations of the experimental $C_{p,m}^{pg}$ from the calculated values by AGA8-DC92 EoS [13,14] are half of the relative deviations with respect to GERG-2008 EoS [10,11] and within the model expanded ($k = 2$) uncertainties. Average absolute relative deviations are $\Delta_{AAD}$ (AGA8-DC92) = 0.10 % and $\Delta_{AAD}$ (GERG-2008) = 0.22 %.

$C_{p,m}^{pg}$ results are fitted to Planck-Einstein-like functions as those derived from statistical mechanics for the shape of the heat capacity as function of the temperature:

$$\frac{C_{p,m}^{pg}}{R} = v_0 + v_1 \frac{\left(u_1/T\right)^2 e^{u_1/T}}{\left(e^{u_1/T} - 1\right)^2} \tag{9}$$

with $v_0 = 4.856 \pm 0.049$, $v_1 = 6.91 \pm 0.14$, and $u_1 = 1420 \pm 28$ K for the (0.60 $CO_2$ + 0.40 $C_3H_8$) mixture and $v_0 = 4.349 \pm 0.021$, $v_1 = 4.818 \pm 0.051$, and $u_1 = 1402 \pm 16$ K for the (0.80 $CO_2$ + 0.20 $C_3H_8$) mixture, where the terms are empirical coefficients. The estimated expanded ($k = 2$) experimental uncertainties $U(C_{p,m}^{pg})$ where used as weights in the regression, with root mean squares of the residuals $\Delta_{RMS}$ equal to 0.043 % and 0.020 %, within the average $U(C_{p,m}^{pg}) = 0.28$ % and 0.15 % for the (0.60 $CO_2$ + 0.40 $C_3H_8$) and (0.80 $CO_2$ + 0.20 $C_3H_8$) mixtures, respectively.

### 4.3 Virial coefficients.

Second acoustic virial coefficient $\beta_a$ is obtained, for each isotherm, from $A_0$ and $A_1$ of Table 7 by:

$$\beta_a = \frac{M}{\gamma^{pg}} A_1 = RT \frac{A_1}{A_0} \tag{10}$$

whereas third acoustic virial coefficient $\gamma_a$ is derived, at each temperature, from $A_0$, $A_1$, and $A_2$ fitted parameters of Table 7:

$$\gamma_a = \frac{RTM}{\gamma^{pg}} A_2 + B\beta_a = \frac{RT}{A_0}\left(RTA_2 + BA_1\right) \quad (11)$$

where $B$ is second density virial coefficient of the mixture for each isotherm. Table 8 lists the results of $\beta_a$ and $\gamma_a$, together with their relative expanded ($k = 2$) uncertainty $U(\beta_a)$ between (0.33 to 1.1) % and $U(\gamma_a)$ ranging from (2.4 to 35) %, and the comparison with the predicted values using the AGA8-DC92 EoS [13,14] and GERG-2008 EoS [10,11]. Relative deviations of $\beta_a$ from EoS could only be explained within the uncertainty for the GERG-2008 EoS [10,11], in particular, at the intermediate isotherms of mixture (0.80 $CO_2$ + 0.20 $C_3H_8$); $\Delta_{AAD}$ (GERG) = 0.89 %, for the two mixtures, is slightly larger than the average $U(\beta_a)$ = 0.68 %. In contrast, discrepancies with the AGA8-DC92 [13,14] are always negative and around six times the mean $U(\beta_a)$, with a $\Delta_{AAD}$ (AGA) = 4.3 %. For $\gamma_a$, there is no good agreement with any of the two studied models, with $\Delta_{AAD}$ (GERG) = 15 % and $\Delta_{AAD}$ (AGA) = 23 % in comparison with the mean $U(\gamma_a)$ = 12 %, although our results are more in line with the predicted values using GERG-2008 EoS [10,11]. The relative differences tend to decrease with increasing temperature, but with no clear trend with composition. This behavior for the two acoustic virial coefficients is expected since, as discussed above with regard to speed of sound data, the overall relative deviations are lower for the higher isotherms and more consistent with the GERG-2008 model [10,11] than the AGA8-DC92 model [13,14].

For further validation of our speed of sound data, second density virial coefficients $B$ of the ($CO_2$ + $C_3H_8$) mixture were derived from the correlations of the perfect-gas heat capacity $C_{p,m}^{pg}$ given at Equation (9) and the estimated second acoustic virial coefficients $\beta_a$ listed in Table 8 by a non-linear regression fitting to two different effective intermolecular potentials $U(r)$, as described in detail in [31]. As $B$ determination is not initially very sensitive to the intermolecular potential used, two common spherical symmetric effective potentials for non-polar molecules were chosen, such as, the hard-core square well (HCSW):

$$\begin{aligned} U(r) &= \infty & r &< \sigma_{SW} \\ U(r) &= -\varepsilon_{SW} & \sigma_{SW} &\leq r \leq g\sigma_{SW} \\ U(r) &= 0 & r &> g\sigma_{SW} \end{aligned} \quad (6)$$

and the Lennard-Jones (12,6) (LJ (12,6)):

$$U(r) = 4\varepsilon_{LJ}\left[\left(\frac{\sigma_{LJ}}{r}\right)^{12} - \left(\frac{\sigma_{LJ}}{r}\right)^{6}\right] \quad (7)$$

whose integration to derive $B$ can be expressed as analytical close form in terms of modified Bessel functions of the first kind [48]. This procedure should be sufficient to perform the regression within our experimental expanded ($k = 2$) uncertainty $U(\beta_a)$. Here, $\sigma_{Sw}$ is the hard-core length, $\varepsilon_{Sw}$ is the square well depth, and $g$ is $\sigma_{Sw}$ times the length of the square well, whereas $\sigma_{LJ}$ is the separation at which $U(r) = -\varepsilon_{LJ}$ and $\varepsilon_{LJ}$ is the depth of the potential well.

The fitting parameters of the regression to the HCSW and LJ (12,6) effective potentials are listed in Table 9 with the $\Delta_{RMS}$ of the residuals of each fitting.

**Table 9.** Regression parameters of the hard-core square well (HCSW) and Lennard-Jones (LJ (12,6)) effective intermolecular potentials $U(r)$ and the root mean square ($\Delta_{RMS}$) of the residuals of the fitting. $\sigma_{SW}$ is the hard-core length, $\varepsilon_{SW}$ is the well depth, $g_{SW}$ is $\sigma_{SW}$ times the length of the square well. $\sigma_{LJ}$ is the separation at which $U(r) = -\varepsilon_{LJ}$ and $\varepsilon_{LJ}$ is the depth of the potential well.

| | HCSW | | | | LJ (12,6) | | |
|---|---|---|---|---|---|---|---|
| $\sigma_{SW}$ / Å | $g_{SW}$ | $\varepsilon_{SW}$ / eV | $\Delta_{RMS}$ of residuals / % | | $\sigma_{LJ}$ / Å | $\varepsilon_{LJ}$ / eV | $\Delta_{RMS}$ of residuals / % |
| (0.60 $CO_2$ + 0.40 $C_3H_8$) | | | | | | | |
| 41.821 | 1.5803 | 0.018227 | 0.27 | | 4.9112 | 0.017492 | 0.49 |
| (0.80 $CO_2$ + 0.20 $C_3H_8$) | | | | | | | |
| 33.315 | 1.3695 | 0.033674 | 0.40 | | 4.5438 | 0.017395 | 1.5 |

Results of $B_{exp}$, whose estimated mean expanded ($k = 2$) uncertainty is $U(B_{exp}) = 1.1$ % using the Monte Carlo method [44], are compared to predictions using the GERG-2008 EoS [10,11] in Figure 8 and Table 10. Hereafter, the comparison with the AGA8-DC92 EoS [13,14] is discarded due to the high disagreement with this model, with relative deviations of 9.0 % at best.

**Table 10.** Second density virial coefficient $B$ and interaction second density virial coefficient $B_{12}$ at the experimental temperatures from the deductions of the hard-core square well (HCSW) and Lennard-Jones (LJ (12,6)) effective intermolecular potentials $U(r)$ for the ($CO_2$ + $C_3H_8$) mixtures, together with their corresponding relative expanded ($k = 2$) uncertainties.

| | (0.60 $CO_2$ + 0.40 $C_3H_8$) | | | (0.80 $CO_2$ + 0.20 $C_3H_8$) | | | Average values by the HCSW potential for the two mixtures | | |
|---|---|---|---|---|---|---|---|---|---|
| $T$ / K | $B_{HCSW}$ / cm$^3 \cdot$mol$^{-1}$ | $B_{LJ}$ / cm$^3 \cdot$mol$^{-1}$ | $10^2 \cdot U(B)$ | $B_{HCSW}$ / cm$^3 \cdot$mol$^{-1}$ | $B_{LJ}$ / cm$^3 \cdot$mol$^{-1}$ | $10^2 \cdot U(B)$ | $B_{12}$ / cm$^3 \cdot$mol$^{-1}$ | $10^2 \cdot U(B_{12})$ | $10^2 \cdot \Delta B_{12,GERG}$ |
| 273.16 | −225.5 | −221.8 | 1.1 | −186.1 | −173.8 | 0.54 | −211.5 | | −1.36 |
| 300.00 | −186.1 | −184.8 | 1.3 | −149.3 | −144.7 | 0.60 | −172.5 | | −0.57 |
| 325.00 | −157.0 | −156.9 | 1.4 | −123.7 | −122.8 | 0.66 | −144.3 | 2.8 | 0.01 |
| 350.00 | −133.4 | −133.7 | 1.6 | −103.6 | −104.6 | 0.72 | −121.6 | | 0.53 |
| 375.00 | −113.7 | −114.2 | 1.8 | −87.6 | −89.2 | 0.80 | −103.1 | | 1.11 |

As can be seen in Figure 8, our results are more consistent with the GERG-2008 EoS [10,11] for the deductions from the HCSW and for the mixture with lower molar fraction of propane. The absolute average deviations are $\Delta_{AAD}$ (GERG) = (1.1 and 0.5) % for the mixtures (0.60 $CO_2$ + 0.40 $C_3H_8$) and (0.80 $CO_2$ + 0.20 $C_3H_8$), respectively, within $U(B_{exp})$ at all the temperature range studied for the latter mixture. This, along with the fact that the $\Delta_{RMS}$ of the residuals for the fitting from the LJ (12,6) potential are as high as 1.5 % and outside $U(\beta_a)$ for the mixture (0.80 $CO_2$ + 0.20 $C_3H_8$), leads us to neglect the $B_{exp}$ values obtained from this effective potential for the determination of the interaction second density virial coefficients $B_{12}$:

$$B_{12} = \left[ B_{exp} - \left( x_1^2 B_{11} + x_2^2 B_{22} \right) \right] / (2 x_1 x_2) \tag{8}$$

Here, $B_{11}$ is the second density virial coefficient of pure carbon dioxide of mole fraction $x_1$ and $B_{22}$ is the second density virial coefficient of pure propane of mole fraction $x_2$, obtained from the corresponding equations of state [19] for pure carbon dioxide and [20] for pure propane, with estimated expanded ($k = 2$) uncertainties $U(B_{11}) = 0.5$ % and $U(B_{22}) = 2.0$ %, respectively [49].

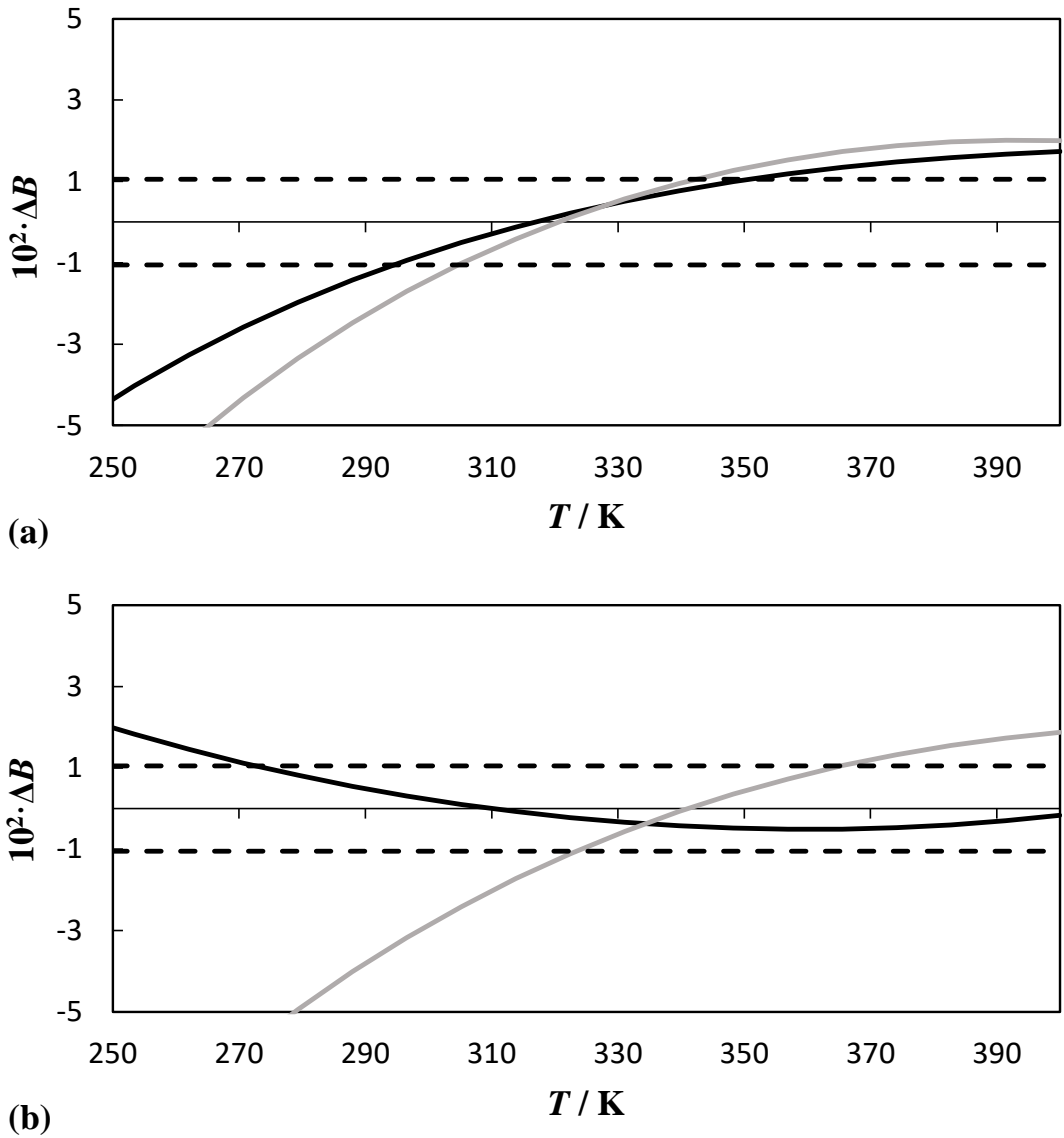

**Figure 8.** Relative deviations of the second density virial coefficient $\Delta B = (B_{exp} - B_{ref})/B_{ref}$ from GERG-2008 EoS [10,11], as a function of temperature for (a) the (0.60 $CO_2$ + 0.40 $C_3H_8$) mixture and (b) the (0.80 $CO_2$ + 0.20 $C_3H_8$) mixture using HCSW (black line) and LJ (12,6) (grey line) effective intermolecular potentials. Dashed lines depict the mean expanded ($k = 2$) uncertainty of this work

The relative deviations of interaction second density virial coefficient, ($B_{12,exp}$ − $B_{12,ref}$)/$B_{12,ref}$, from the GERG-2008 EoS [10,11] and the literature data are depicted in Figure 9.

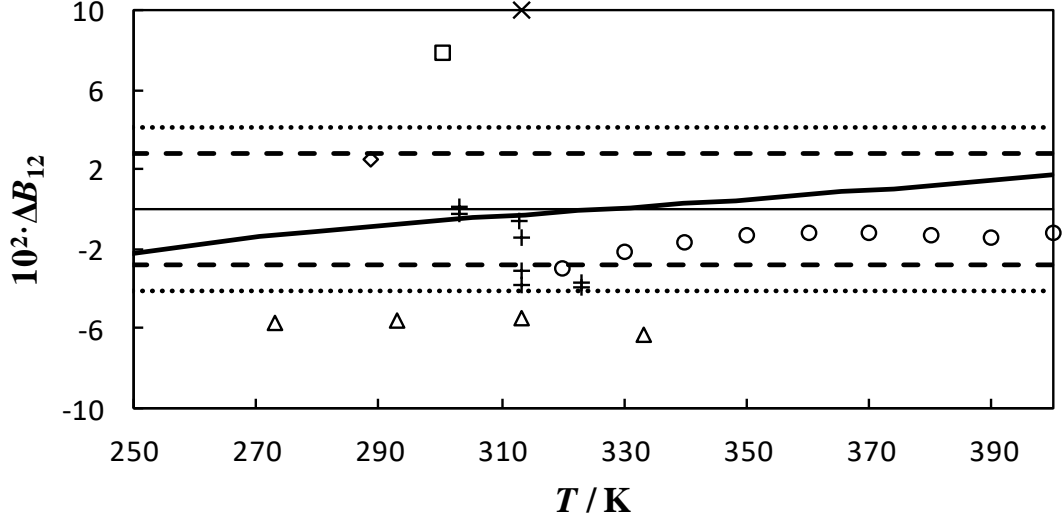

**Figure 9.** Relative deviations of the interaction second density virial coefficient $\Delta B_{12}$ = ($B_{12,exp}$ − $B_{12,ref}$)/$B_{12,ref}$ as function of temperature, with $B_{12,exp}$ obtained from the average of the HCSW deductions for the two ($CO_2$ + $C_3H_8$) mixtures: solid line from the GERG-2008 EoS [10,11], △ Jaeschke et al. [50], ◇ Mason et al. [51], □ Bougard et al. [52], + McElroy et al. [53], × Sie et al. [54], ○ Feng et al. [55]. Dashed lines depict the mean expanded ($k$ = 2) uncertainty of this work and dotted lines depict the average expanded ($k$ = 2) uncertainty of literature.

$B_{12,exp}$ were obtained from the average of the results derived by the HCSW effective intermolecular potential for the two ($CO_2$ + $C_3H_8$) mixtures. The literature data include values reported by Jaeschke et al. [50], Mason et al. [51], Bougard et al. [52], McElroy et al. [53], Sie et al. [54], and Feng et al. [55] which were obtained by different techniques such as densimeters based on the Archimedes principle, expansion techniques or Burnett method, and gas chromatography. Comparison to the GERG-2008 EoS [10,11] is performed to the average predicted values from the EoS for the two studied mixtures because, although $B_{12}$ is only a function of the temperature, the reference model shows a slight dependence with the composition. The estimated expanded ($k = 2$) uncertainty $U(B_{12,\text{exp}}) = 2.8$ %, for the experimental values is similar to the mean of the literature values $U(B_{12,\text{Literature}}) = 4.1$ %. As depicted in Figure 9, the discrepancies of our results $B_{12,\text{exp}}$ from the GERG-2008 EoS [10,11] are explained within $U(B_{12,\text{exp}})$, while the values of Sie et al. [54] and Bougard et al [52] are not in agreement with ours neither with the model nor with the other authors. In addition, the values of Mason et al. [51] and McElroy et al. [53] are in agreement within the mutual uncertainty of the literature and the experimental one with our estimations. Furthermore, the most recent results of Feng et al. [55] show a nearly constant discrepancy with our values in all the temperature range, but within the limit of the experimental uncertainty. Finally, the data set of Jaeschke et al. [50] deviates by about −6 % for all the isotherms. The overall absolute average deviation of the experimental $B_{12}$ of this work from the literature data, discarding the outlier values of Sie et al. [54] and Bougard et al [52], is $\Delta_{\text{AAD}} = 2.4$ % which is reasonable in comparison to its uncertainty.

**4.4 Phase Equilibria.**

Experimental phase equilibria data are compared to the predicted values of GERG-2008 equation [10,11] as is listed in Tables 5 and 6 shown above and they are also depicted in Figure 10.

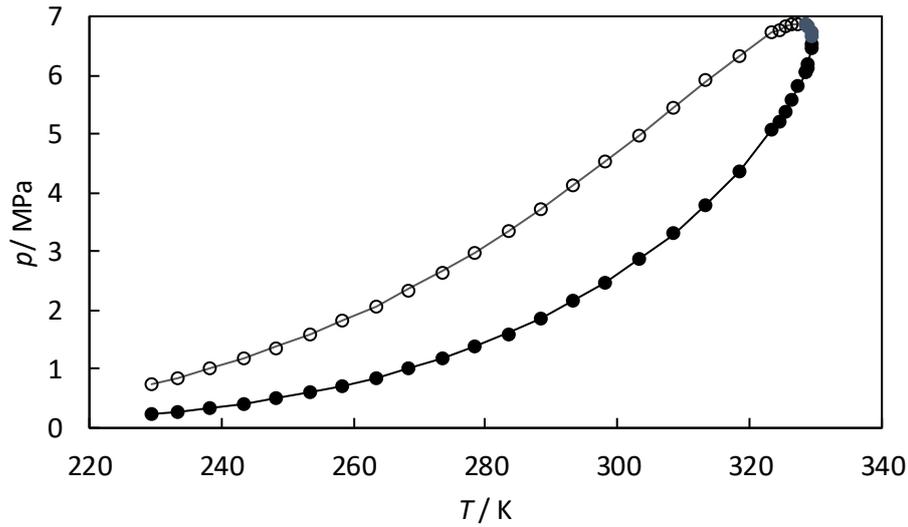

(a)

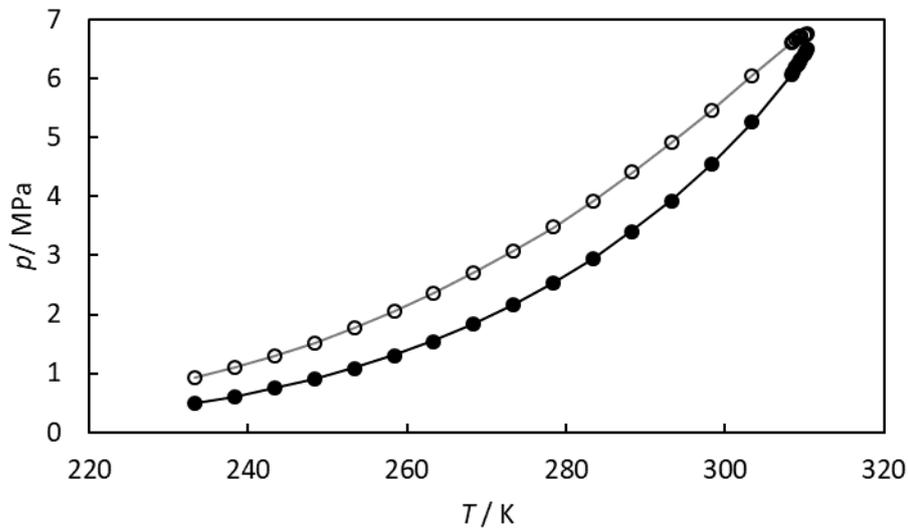

(b)

**Figure 10**. Phase envelope: a) the (0.60 CO$_2$ + 0.40 C$_3$H$_8$) mixture b) the (0.80 CO$_2$ + 0.80 C$_3$H$_8$) mixture. (o) bubble points; (◉) retrograde dew points; (●) dew points; (–) calculated bubble points using GERG-2008 EoS [11]; (–) calculated dew points using GERG-2008 EoS [11].

The average absolute relative pressure deviations are 0.19% and 0.23% for bubble points and dew points of the (0.60 CO$_2$ + 0.40 C$_3$H$_8$) mixture, respectively, and 0.28% and 0.22% for bubble points and dew points of the (0.80 CO$_2$ + 0.20 C$_3$H$_8$) mixture, respectively. All these values are below the equation of state uncertainty which is estimated between 1% to 3% [11]. As seen, the (0.60 $CO_2$ + 0.40 $C_3H_8$) mixture presents a retrograde condensation zone, according to the critical point (327.71 K, 6.8724 MPa) calculated with the GERG-2008 equation, which has also been determined experimentally.

In addition, experimental results are also compared with the scarce literature data [61-64]. These data were selected between those whose compositions are close to our mixtures. Figure 11 presents the relative deviations between the different sets of experimental pressures and the corresponding values calculated using the GERG-2008 EoS [10,11].

This figure clearly shows that our deviations are much lower than the deviations obtained for the literature data, and our results scatter around the zero value and do not show any tendency.

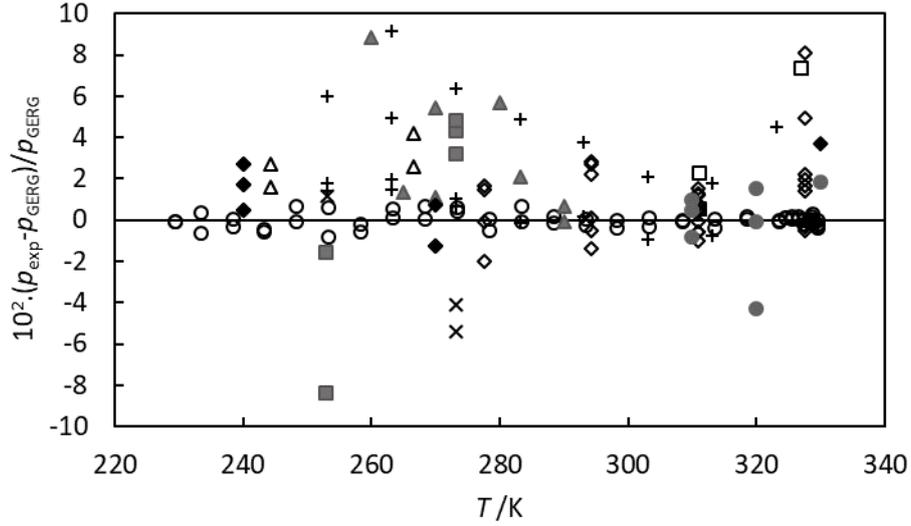

(a)

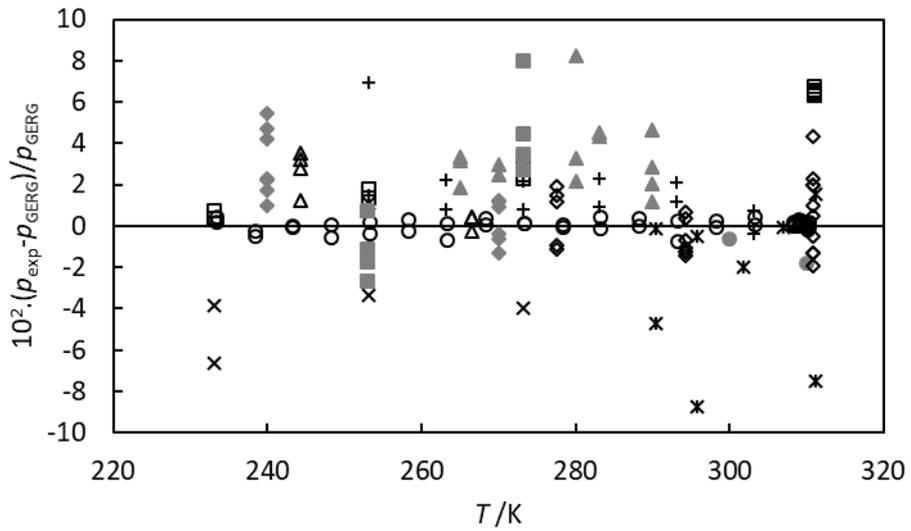

(b)

**Figure 11**. Relative deviations of pressure for ($CO_2$ + $C_3H_8$) mixtures: (a) $x_{CO2}$≈0.6; (b) $x_{CO2}$≈0.8. Symbols: (o) this work; (×) Akers et al. [61]; (Δ) Hamman et al. [62]; (□) Niesen et al. [63]; (+) Kim et al. [64]; (◊) Reamer et al. [65]; (■) Nagahama et al. [66]; (♦) Yucelen et al. [67]; (▲) Tanaka et al. [68]; (●) Nagata et al. [69]; (✶)Poettmann et al. [70].

The average absolute relative deviations obtained for the $x_{CO2}$≈0.6 ($CO_2$ + $C_3H_8$) mixture are: 3.91% for data from [61], 2.77% for [62], 4.63% for [63], 2.83% for [64], 1.53% for [65], 4.48% for [66], 1.67% for [67], 3.93% for [68], 1.43% for [69]. Lower deviations (in general) were found for the $x_{CO2} \approx 0.8$ ($CO_2 + C_3H_8$) mixture: 3.78% for [61], 1.70% for [62], 2.27% for [63], 1.83% for [64], 1.38% for [65], 2.95% for [66], 2.09% for [67], 3.95% for [68], 1.21% for [69], 4.72% for [70]. These values are mostly in agreement with the uncertainty of the EoS.

One of the main advantages of our technique is that the composition is fixed, and its uncertainty is lower than other techniques where the composition is measured. As is indicated in [11], the main contribution to phase equilibria uncertainty is due to the composition uncertainty.

## 5. Conclusions.

New high-accurate experimental speed of sound and phase equilibria data for two binary mixtures of carbon dioxide and propane, with nominal compositions of (0.60 $CO_2$ + 0.40 $C_3H_8$) and (0.80 $CO_2$ + 0.20 $C_3H_8$) are reported. Speed of sound measurements were performed with an acoustic stainless-steel spherical resonator achieving an expanded ($k = 2$) uncertainty of 250 parts in $10^6$ (0.025 %).

From these data, perfect-gas heat capacity ratios and heat capacities were obtained with an expanded ($k = 2$) uncertainty between (0.15 to 0.28) %, second acoustic virial coefficients were deduced with an expanded ($k = 2$) uncertainty of (0.33 to 1.1) %, and third acoustic virial coefficients were determined with an expanded ($k = 2$) uncertainty of (2.4 to 35) %. Speed of sound results are consistent within the large stated 1% uncertainty of reference models AGA8-DC92 Eos [13,14] and GERG-2008 EoS [10,11], with deviations that increase towards higher pressures, showing a better agreement in the order of 0.06 % according to the latter equation of state at high pressures. On the contrary, the comparison of the derived heat capacities in the limit of zero pressure is more satisfactory with respect to the AGA8-DC92 EoS [13,14], yielding deviations less than 0.1 %.

The low uncertainty of the acoustic virial coefficients allows us to analyze our data using the effective intermolecular potentials of hard-core square well and Lennard-Jones (12,6) in order to derive second density virial coefficients and the corresponding interaction coefficients for this binary mixture. The relative differences of the interaction second density virial coefficient are consistent regarding to the GERG-2008 EoS [10,11] and within the claimed uncertainty of most authors in the literature, with deviations less than 5 % in comparison with an experimental expanded ($k = 2$) uncertainty of 2.8 %.

Finally, phase equilibria data, measured using a cylindrical microwave resonator, allow accurate bubble or dew points with expanded uncertainties ($k = 2$) (40 mK and $4.0 \cdot 10^{-4}$ (p/Pa) + 1400 Pa). The deviations obtained from GERG-2008 EoS [10,11] are lower than the uncertainty of the EoS.

This work tries to enrich the set of accurate data and to fill the gap that exists for binary mixtures composed of carbon dioxide and propane, which will benefit the development of improved correlations for the reference equations of state and, hence, the calculations made during the design and control stages of the mentioned process involving these mixtures, where precise and accurate information of the thermodynamic properties is required for the understanding of the whole system.

**Acknowledgements.**

This work was supported by FEDER/Ministerio de Ciencia, Innovación y Universidades – Agencia Estatal de Investigación (Project ENE2017-88474-R) and FEDER/Junta de Castilla y León (Project VA280P18).

**References.**

[1] Niu B, Zhang Y. Experimental study of the refrigeration cycle performance for the R744/R290 mixtures. Int J Refrig 2007;30:37–42. https://doi.org/10.1016/j.ijrefrig.2006.06.002.

[2] Kim JH, Cho JM, Kim MS. Cooling performance of several CO2/propane mixtures and glide matching with secondary heat transfer fluid. Int J Refrig 2008;31:800–6. https://doi.org/10.1016/j.ijrefrig.2007.11.009.

[3] Garg P, Kumar P, Srinivasan K, Dutta P. Evaluation of carbon dioxide blends with isopentane and propane as working fluids for organic Rankine cycles. Appl Therm Eng 2013;52:439–48. https://doi.org/10.1016/j.applthermaleng.2012.11.032.

[4] Feng L, Zheng D, Chen J, Dai X, Shi L. Exploration and Analysis of CO2 + Hydrocarbons Mixtures as Working Fluids for Trans-critical ORC. Energy Procedia 2017;129:145–51. https://doi.org/10.1016/j.egypro.2017.09.191.

[5] Illés V, Szalai O, Then M, Daood H, Perneczki S. Extraction of hiprose fruit by supercritical CO2 and propane. J Supercrit Fluids 1997;10:209–18. https://doi.org/10.1016/S0896-8446(97)00018-1.

[6] King JW, Holliday RL, List GR, Snyder JM. Hydrogenation of vegetable oils using mixtures of supercritical carbon dioxide and hydrogen. JAOCS, J Am Oil Chem Soc 2001;78:107–13. https://doi.org/10.1007/s11746-001-0229-8.

[7] Palla C, Hegel P, Pereda S, Bottini S. Extraction of jojoba oil with liquid CO2 + propane solvent mixtures. J Supercrit Fluids 2014;91:37–45. https://doi.org/10.1016/j.supflu.2014.04.005.

[8] Correa M, Mesomo MC, Pianoski KE, Torres YR, Corazza ML. Extraction of inflorescences of Musa paradisiaca L. using supercritical CO2 and compressed

propane. J Supercrit Fluids 2016;113:128–35. https://doi.org/10.1016/j.supflu.2016.03.016.

[9] Luo P, Zhang Y, Wang X, Huang S. Propane-enriched CO 2 immiscible flooding for improved heavy oil recovery. Energy and Fuels 2012;26:2124–35. https://doi.org/10.1021/ef201653u.

[10] Kunz O, Klimeck R, Wagner W, Jaeschke M. GERG Technical Monograph 15 The GERG-2004 wide-range equation of state for natural gases and other mixtures. Düsseldorf: 2007.

[11] Kunz O, Wagner W. The GERG-2008 Wide-Range Equation of State for Natural Gases and Other Mixtures: An Expansion of GERG-2004. J Chem Eng Data 2012;57:3032–91. https://doi.org/10.1021/je300655b.

[12] Lin CW, Trusler JPM. Speed of sound in (carbon dioxide + propane) and derived sound speed of pure carbon dioxide at temperatures between (248 and 373) K and at pressures up to 200 MPa. J Chem Eng Data 2014;59:4099–109. https://doi.org/10.1021/je5007407.

[13] Transmission Measurement Committee. AGA Report No. 8 Part 2 Thermodynamic Properties of Natural Gas and Related Gases GERG–2008 Equation of State. 2017.

[14] International Organization for Standardization. ISO 20765-1 Natural gas — Calculation of thermodynamic properties — Part 1: Gas phase properties for transmission and distribution applications. Genève: 2005.

[15] Susial R, Gómez-Hernández Á, Lozano-Martín D, del Campo D, Martín MC, Segovia JJ. A novel technique based in a cylindrical microwave resonator for high pressure phase equilibrium determination. J Chem Thermodyn 2019;135. https://doi.org/10.1016/j.jct.2019.03.027.

[16]   International Organization for Standardization. ISO 6142-1 Gas analysis — Preparation of calibration gas mixtures — Part 1: Gravimetric method for Class I mixtures. Genève: 2014.

[17]   International Organization for Standardization. Gas analysis - Comparison methods for determining and checking the compposition of calibration gas mixtures. 2006.

[18]   Lemmon EW, Bell IH, Huber ML, McLinden MO. NIST Standard Reference Database 23: Reference Fluid Thermodynamic and Transport Properties-REFPROP, Version 10.0, National Institute of Standards and Technology 2018:135. https://doi.org/https://doi.org/10.18434/T4/1502528.

[19]   Span R, Wagner W. A new equation of state for carbon dioxide covering the fluid region from the triple-point temperature to 1100 K at pressures up to 800 MPa. J Phys Chem Ref Data 1996;25:1509–96. https://doi.org/10.1063/1.555991.

[20]   Lemmon EW, McLinden MO, Wagner W. Thermodynamic properties of propane. III. A reference equation of state for temperatures from the melting line to 650 K and pressures up to 1000 MPa. J Chem Eng Data 2009;54:3141–80. https://doi.org/10.1021/je900217v.

[21]   Pérez-Sanz FJ, Segovia JJ, Martín MC, Del Campo D, Villamañán MA. Speeds of sound in (0.95 N2 + 0.05 CO and 0.9 N2 + 0.1 CO) gas mixtures at T = (273 and 325) K and pressure up to 10 MPa. J Chem Thermodyn 2014;79:224–9. https://doi.org/10.1016/j.jct.2014.07.022.

[22]   Lozano-Martín D, Rojo A, Martín MC, Vega-Maza D, Segovia JJ. Speeds of sound for (CH4 + He) mixtures from p = (0.5 to 20) MPa at T = (273.16 to 375) K. J Chem Thermodyn 2019;139. https://doi.org/10.1016/j.jct.2019.07.011.

[23]   Mehl JB. Analysis of resonance standing-wave measurements. J Acoust Soc Am

1978;64:1523–5. https://doi.org/10.1121/1.382096.

[24] Ewing MB, Trusler JPM. On the analysis of acoustic resonance measurement. J Acoust Soc Am 1989;85:1780–2. https://doi.org/10.1121/1.397970.

[25] Pérez-Sanz FJ, Martín MC, Chamorro CR, Fernández-Vicente T, Segovia JJ. Heat capacities and acoustic virial coefficients for a synthetic coal mine methane mixture by speed of sound measurements at T = (273.16 and 250.00) K. J Chem Thermodyn 2016;97:137–41. https://doi.org/10.1016/j.jct.2016.01.020.

[26] Lozano-Martín D, Segovia JJ, Martín MC, Fernández-Vicente T, del Campo D. Speeds of sound for a biogas mixture $CH_4 + N_2 + CO_2 + CO$ from p = (1–12) MPa at T = (273, 300 and 325) K measured with a spherical resonator. J Chem Thermodyn 2016;102:348–56. https://doi.org/10.1016/j.jct.2016.07.033.

[27] Segovia JJ, Lozano-Martín D, Martín MC, Chamorro CR, Villamañán MA, Pérez E, et al. Updated determination of the molar gas constant R by acoustic measurements in argon at UVa-CEM. Metrologia 2017;54:663–73. https://doi.org/10.1088/1681-7575/aa7c47.

[28] Preston-Thomas H. The International temperature scale of 1990 (ITS-90). Metrologia 1990;27:3–10. https://doi.org/10.1088/0026-1394/27/1/002.

[29] Preston-Thomas H. The International Temperature Scale of 1990 (ITS-90). Metrologia 1990;27:107–107. https://doi.org/10.1088/0026-1394/27/2/010.

[30] Mehl JB, Moldover MR. Precondensation phenomena in acoustic measurements. J Chem Phys 1982;77:455–65. https://doi.org/10.1063/1.443627.

[31] Lozano-Martín D, Martín MC, Chamorro CR, Tuma D, Segovia JJ. Speed of sound for three binary ($CH_4 + H_2$) mixtures from p = (0.5 up to 20) MPa at T = (273.16 to 375) K. Int J Hydrogen Energy 2020;45:4765–83. https://doi.org/10.1016/j.ijhydene.2019.12.012.

[32] Estrada-Alexanders AF, Trusler JPM. Speed of sound in carbon dioxide at temperatures between (220 and 450) K and pressures up to 14 MPa. J Chem Thermodyn 1998;30:1589–601. https://doi.org/10.1006/jcht.1998.0428.

[33] Estrada-Alexanders AF, Hurly JJ. Kinematic viscosity and speed of sound in gaseous CO, CO2, SiF4, SF6, C4F8, and NH3 from 220 K to 375 K and pressures up to 3.4 MPa. J Chem Thermodyn 2008;40:193–202. https://doi.org/10.1016/j.jct.2007.07.002.

[34] Liu Q, Feng X, An B, Duan Y. Speed of sound measurements using a cylindrical resonator for gaseous carbon dioxide and propene. J Chem Eng Data 2014;59:2788–98. https://doi.org/10.1021/je500424b.

[35] Trusler JPM, Zarari MP. The speed of sound in gaseous propane at temperatures between 225 K and 375 K and at pressures up to 0.8 MPa. J Chem Thermodyn 1996;28:329–35. https://doi.org/10.1006/jcht.1996.0032.

[36] Hurly JJ, Gillis KA, Mehl JB, Moldover MR. The viscosity of seven gases measured with a Greenspan viscometer. Int J Thermophys 2003;24:1441–74. https://doi.org/10.1023/B:IJOT.0000004088.04964.4c.

[37] Meier K, Kabelac S. Thermodynamic properties of propane. IV. Speed of sound in the liquid and supercritical regions. J Chem Eng Data 2012;57:3391–8. https://doi.org/10.1021/je300466a.

[38] Stephenson JC, Wood RE, Moore CB. Vibrational Relaxation of Laser-Excited CO 2 -Polyatomic Mixtures 2003;4813:1–5.

[39] Estela-Uribe JF, Trusler JPM, Chamorro CR, Segovia JJ, Martín MC, Villamañán MA. Speeds of sound in {(1 - x)CH4 + xN2} with x = (0.10001, 0.19999, and 0.5422) at temperatures between 170 K and 400 K and pressures up to 30 MPa. J Chem Thermodyn 2006;38:929–37.

https://doi.org/10.1016/j.jct.2005.10.006.

[40]  Herzberg G. Molecular Spectra and Molecular Structure II. Infrared and Raman Spectra of polyatomic molecules. Toronto, New York, London: D. Van Nostrand Company; 1945.

[41]  Shimanouchi T. Tables of Molecular Vibrational Frequences Consolidated, vol. 1. Natl Bur Stand 1972:1–16. https://doi.org/10.6028/NBS.NSRDS.39.

[42]  Martin E, Hernandez A, Sanchez MC, Zamarro JM, Margineda J. Automatic measurement of Q factor and resonant frequency of microwave resonators. J Phys E 1981;14:961–2. https://doi.org/10.1088/0022-3735/14/8/016.

[43]  JCGM. Evaluation of measurement data — Guide to the expression of uncertainty in measurement. 2008.

[44]  Joint Committee for Guides in Metrology. Evaluation of measurement data — Supplement 1 to the "Guide to the expression of uncertainty in measurement" — Propagation of distributions using a Monte Carlo method. vol. JCGM 101:2. 2008.

[45]  Fischer J, Fellmuth B, Gaiser C, Zandt T, Pitre L, Sparasci F, et al. The Boltzmann project. Metrologia 2018;55:R1–20. https://doi.org/10.1088/1681-7575/aaa790.

[46]  Jaeschke M, Schley P. Ideal-gas thermodynamic properties for natural-gas applications. Int J Thermophys 1995;16:1381–92. https://doi.org/10.1007/BF02083547.

[47]  Chao J, Wilhoit RC, Zwolinski BJ. Ideal Gas Thermodynamic Properties of Ethane and Propane. J Phys Chem Ref Data 1973;2:427–38. https://doi.org/10.1063/1.3253123.

[48]  Vargas P, Muñoz E, Rodriguez L. Second virial coefficient for the Lennard–

Jones potential. Phys A Stat Mech Its Appl 2001;290:92–100. https://doi.org/10.1016/S0378-4371(00)00362-9.

[49] Dymond JH, Marsh KN, Wilhoit RC. Virial Coefficients of Pure Gases and Mixtures - Subvolume B. vol. 21B. Berlin/Heidelberg: Springer-Verlag; 2002. https://doi.org/10.1007/b71692.

[50] Jaeschke. M, Audibert. S, van Canegham. P, Humphreys. AE, Janssen-van Rosemalen. R, Pellei. Q, et al. GERG Tech. Monogr. TM2 - High Accuracy Compressibility factor Calculation for Natural Gases and Similar Mixtures by Use of a Truncated Virial Equation. Düsseldorf: 1989.

[51] Mason DMA, Eakin BE. Compressibility Factor of Fuel Gases at 60° F. and 1 Atm. J Chem Eng Data 1961;6:499–504. https://doi.org/10.1021/je60011a006.

[52] Bougard J, Jadot R. Second coefficient du viriel de mélanges binaires d'halocarbones. J Chim Phys 1976;73:415–7. https://doi.org/10.1051/jcp/1976730415.

[53] McElroy PJ, Kee LL, Renner CA. Excess Second Virial Coefficients for Binary Mixtures of Carbon Dioxide with Methane, Ethane, and Propane. J Chem Eng Data 1990;35:314–7. https://doi.org/10.1021/je00061a024.

[54] Sie ST, Van Beersum W, Rijnders GWA. High-Pressure Gas Chromatography and Chromatography with Supercritical Fluids. I. The Effect of Pressure on Partition Coefficients in Gas-Liquid Chromatography with Carbon Dioxide as a Carrier Gas. Sep Sci 1966;1:459–90. https://doi.org/10.1080/01496396608049460.

[55] Feng XJ, Liu Q, Zhou MX, Duan YY. Gaseous pvTx properties of mixtures of carbon dioxide and propane with the burnett isochoric method. J Chem Eng Data 2010;55:3400–9. https://doi.org/10.1021/je100148h.